\documentclass[12pt,preprint]{aastex}

\usepackage{multirow}

\newcommand{\rosat}{{\it ROSAT}}
\newcommand{\pspc}{\rosat\ PSPC}

\newcommand{\twomass}{2MASS}
\newcommand{\usno}{\mbox{USNO-A2}}
\newcommand{\bsc}{RASS/BSC}
\newcommand{\psc}{2MASS/PSC}
\newcommand{\xsc}{2MASS/XSC}
\newcommand{\simbad}{SIMBAD}
\newcommand{\xidp}{\citetext{R00}}
\newcommand{\xidt}{R00}

\newcommand{\eg}{{e.g.}}
\newcommand{\ie}{{i.e.}}
\newcommand{\cf}{{cf.}}
\newcommand{\xray}{\mbox{X-ray}}
\newcommand{\ir}{NIR}

\newcommand{\ee}[1]{\mbox{$10^{#1}$}}
\newcommand{\tee}[1]{\mbox{$\times 10^{#1}$}}

\shorttitle{XID II: \bsc--\psc\ cross-association}
\shortauthors{C.~B. Haakonsen \& R.~E. Rutledge}

\begin{document}

\title{XID II: Statistical Cross-Association of \rosat\ Bright Source
Catalog X-ray Sources with \twomass\ Point Source Catalog
Near-Infrared Sources}
\author{Christian Bernt Haakonsen \& Robert E. Rutledge}
\affil{Physics Department, McGill University, 3600 rue University, Montreal 
QC H3A 2T8, Canada; christian.haakonsen@mail.mcgill.ca, rutledge@physics.mcgill.ca}

\begin{abstract}
The 18806 \rosat\ All Sky Survey Bright Source Catalog (\bsc) \xray\
sources are quantitatively cross-associated with near-infrared (\ir)
sources from the Two Micron All Sky Survey Point Source Catalog
(\psc). An association catalog is presented, listing the most likely
counterpart for each \bsc\ source, the probability $P_{\rm id}$ that
the \ir\ source and \xray\ source are uniquely associated, and the probability
$P_{\rm no-id}$ that none of the \psc\ sources are associated with the
\xray\ source. The catalog includes 3853 high quality ($P_{\rm
id}>0.98$) \xray--\ir\ matches, 2280 medium quality ($0.98\geq P_{\rm
id}>0.9$) matches, and 4153 low quality ($0.9\geq P_{\rm id}>0.5$)
matches. Of the high quality matches, 1418 are associations 
that are not listed in the \simbad\ database, and 
for which no high quality match with a \usno\ optical source 
was presented for the \bsc\ source in previous work. 
The present work 
offers a significant number of new associations with \bsc\ objects 
that will require optical/\ir\ spectroscopy for classification.  For
example, of the 6133 $P_{\rm id}>0.9$ \psc\ counterparts presented
in the association catalog, 2411 have no classification listed in the \simbad\
database. These \psc\ sources will likely include scientifically
useful examples of known source classes of \xray\ emitters (white
dwarfs, coronally active stars, active galactic nuclei), but may also
contain previously unknown source classes.
It is determined that all coronally active stars in the \bsc\ should 
have a counterpart in the \psc, and that the unique association 
of these \bsc\ sources with their \ir\ counterparts thus is confusion limited.
\end{abstract}


\maketitle

\clearpage
\section{INTRODUCTION}
\label{sec:intro}

The \rosat\ All Sky Survey Bright Source Catalog \citep[\bsc ;][]{voges99} contains flux
and positional information for 18806 X-ray sources\footnote{The 
reference \citep{voges99} refers to 18811 sources, but 5 duplicates were removed in the 
most recent version of the catalog. See 
\url{ftp://ftp.xray.mpe.mpg.de/rosat/catalogues/rass-bsc/1.3\_vs\_1.4.diff}
for details (accessed 2009 May 5).}, down to a limiting 0.1--2.4 keV
count rate of $0.05\ {\rm counts\, s}^{-1}$.  The observations were
made in 1990/91 with the \rosat/PSPC satellite, and at a brightness
limit of $0.1\ {\rm counts\, s}^{-1}$ (5843 sources) the \bsc\
represents a sky coverage of 92\%. The
median positional uncertainty ($1\sigma$, including 6\arcsec\ systematic 
uncertainty) is $11\arcsec$.

The sources in the \bsc\ are the brightest \xray\ sources in the sky.
Brighter sources generally allow for better \xray\ spectra and time
variability observations, so the \bsc\ sources are ideal for studying
the \xray\ properties of their respective classes with modern \xray\
telescopes. However, as of 2008 Aug 1 only 1982 \bsc\ sources have been observed with
either {\em Chandra} or {\em XMM-Newton}. Further, around two 
thirds of the sources in the \bsc\ remain 
unclassified (\cf\ \S\ref{sec:prevclass}), so the number of known
source classes in the \bsc\ and the number of sources of each class is
limited.  Classifying \bsc\ sources therefore not only expands our
knowledge of the brightest \xray\ sources in the sky, but also
provides ideal targets for follow-up observations that can improve our
understanding of specific source classes.  As a larger fraction of the
\bsc\ is classified, the catalog will also become a useful tool in
producing population constraints for various classes of objects, and
the sources that remain unclassified can be targeted in searches for
previously unknown source classes and objects with unusual properties.

To classify \bsc\ sources, one needs spectral and flux measurements in
other wavebands. The error cones of the \xray\ sources often contain
several off-band sources, any one (or none) of which may be a
counterpart. 
If the probability of association between the \xray\ source and one 
or more off-band counterparts is calculated, 
the classification of that \xray--off-band pair/group can be constrained. 
Furthermore, probable off-band counterparts can be studied with ground-based or other
instrumentation, which can provide constraints on the source of X-rays. 
Similarly, the absence of off-band sources near a \bsc\ source
can also constrain its class.  Thus, searching for off-band
counterparts to \bsc\ sources is a good way to advance the
classification and study of these sources.

A summary of previous work identifying likely counterparts for
${>500}$ \bsc\ sources is given in Table
\ref{table:cross-associations}, which also includes the present work
for comparison. Many of these previous cross-associations searched for
counterparts in catalogs of objects of a specific type, like stars
\citep{huensch98, huensch99, makarov00, torres06}, galaxies
\citep{zimmermann01}, or quasars \citep{bade98, zickgraf03}.  Other
previous cross-associations searched for counterparts in catalogs of
sources in one or more specific wave-bands, like optical
\citep{voges99, rutledge00, schwope00, mcglynn04} and/or radio
\citep{bauer00, mcglynn04}.  Only work that specifically addresses
finding counterparts to \bsc\ sources has been included in Table
\ref{table:cross-associations}, while work that includes other \xray\
catalogs without explicitly considering the \bsc\ subset
\citep[e.g.][]{flesch04} has not been included.
\placetable{table:cross-associations}

The \bsc\ sources were quantitatively cross-associated with \usno\ optical sources in 
previous work \citep{rutledge00}. The present work adapts the methods used in that 
\xray--optical cross-association to 
associate \bsc\ sources with Two Micron All Sky Survey Point Source Catalog 
\citep[\psc ;][]{skrutskie06} near-infrared (\ir) sources. The present
work is the first to produce a cross-association between the full
\bsc\ and \psc\ catalogs.

\twomass\ surveyed 99.998\% of the celestial sphere in the near-infrared, 
resulting in a Point Source Catalog containing 470992970 point sources,
and an Extended Source Catalog (XSC) containing 1647599 extended sources. 
The majority of \xsc\ sources are galaxies ($\sim$97\%), while most \psc\ sources 
are stars in the Milky Way. The \psc\ also contains point-source processed 
versions of virtually all the \xsc\ sources, as well as a significant number of 
unresolved galaxies \citep{cutri03}. 
The present work considers only the \psc, for which the 
greatest sensitivity was achieved in the J band ($1.12-1.36\mu m$), with a
$10\sigma$ detection level better than 15.8 mag \citep{skrutskie06}.

This paper is organized as follows.
In \S\ref{sec:method}, the method used 
to perform the statistical cross-association between \bsc\ and \psc\ sources 
is described.
The results of this cross-association are 
discussed in \S\ref{sec:results}, which includes an association catalog containing the most likely \ir\ 
counterpart for each \xray\ source. 
In \S\ref{sec:prevclass}, 
the source types listed in the \simbad\ database for the \psc\ 
sources that appear in the association catalog are discussed.
The present work is compared to 
previous cross-associations in \S\ref{sec:comparison}, 
and discussion and conclusions can be found in \S\ref{sec:conclude}.

\section{ANALYSIS}
\label{sec:method}

The present approach follows that of previous work, in which RASS/BSC X-ray
sources were statistically associated with USNO-A2 optical sources
\citep[\xidt\ hereafter]{rutledge00}.  The main steps to 
this approach are recounted here, as relevant to the present work, while
the earlier work is referred to for a more detailed discussion.

Source fields are defined as regions within $75\arcsec$ of a RASS/BSC
\xray\ source.  For each source field, a number of
non-overlapping, adjacent background fields are extracted (in \xidt,
10 fields were extracted along a line of constant declination through 
the \bsc\ position).

One defines a function, $LR$, which
depends on the joint properties of an X-ray--off-band source
pair, such as the X-ray--off-band source separation $r$ and its
uncertainty, and the off-band magnitude of the candidate counterpart.
$LR$ should be large when the the X-ray source and off-band
counterpart are associated, and small when they are not.

$LR$ is computed for each \xray--off-band source
pair within a source field, and for off-band sources in background
fields as if there is an \xray\ source at the background field
center. 

The reliability of association $R_i$ of an \xray\ source with a given off-band 
source is 
\begin{equation}
\label{eq:R}
R(LR_i) = \frac{n_{\rm true}(LR_i)}{n_{\rm true}(LR_i) + n_{\rm false}(LR_i)}\, ,
\end{equation}
where $LR_i$ is the value of $LR$ for that \xray--off-band source pair, 
$n_{\rm true}(LR_i)$ is the number of true associations (per source field) with that $LR$ value, and 
$n_{\rm false}(LR_i)$ is the number of associations (per source field) with that $LR$ value, where the off-band 
source is an unrelated background source. 
As it is not known {\it a priori} which off-band sources in source fields are unrelated 
background objects and which are true counterparts to a \bsc\ source, 
$R(LR)$ must be approximated based on the number distributions in $LR$ of off-band sources per source 
and background field, $n_{\rm source}(LR)$ and $n_{\rm background}(LR)$ respectively. 
The approximation to $R(LR)$ used is (\cf\ \xidt) 
\begin{equation}
\label{eq:Rapprox}
R(LR) \approx \frac{n_{\rm source}(LR) - n_{\rm background}(LR)}{n_{\rm source}(LR)}\, .
\end{equation}
In the theoretical development below, the formal definition of $R$ (eq.~[\ref{eq:R}]) is used, 
while the approximation to $R$ (eq.~[\ref{eq:Rapprox}]) is used whenever discussing computed values or results.

Assuming that each \xray\ source is either associated uniquely with
one of the $M$ off-band sources in the corresponding source field or
associated with none of them, the probability of unique (exclusive)
association between the X-ray source and the candidate off-band 
counterpart, $P_{{\rm id,} i}$, is calculated:
\begin{equation}
\label{eq:P.id.i.expl}
P_{{\rm id,}i} = \frac{R_i}{(1 - R_i)( 1+\sum_{j=1}^M \frac{R_j}{1 - R_j})}\, .
\end{equation}
Further, the probability that none of the off-band sources
are associated with a given \xray\ source,
\begin{equation}
\label{eq:P.no-id.expl}
P_{\rm no-id} = \frac{1}{1+\sum_{j=1}^M \frac{R_j}{1 - R_j}}\, ,
\end{equation}
is computed. 
This permits the fraction of the \bsc\ that has an 
identifiable counterpart in the off-band catalog (the 'quality') to be determined: 
\begin{equation}
\label{eq:Q}
Q = \langle 1-P_{\rm no-id}\rangle \, ,
\end{equation}
where the average is over all \bsc\ sources. 
The definitions of $P_{\rm id}$ and $P_{\rm no-id}$ given above are the same as in 
the previous work \xidp. The apparent difference is due to the normalization factor 
being written out explicitly in the present work, with common factors being cancelled 
from the numerator and denominator.

The remainder of \S\ref{sec:method} describes the adaptation of the method described above 
to the \bsc--\psc\ cross-association performed in the present work. The main focus of this 
analysis is to identify individual \psc\ sources associated with a \bsc\ source.

\subsection{Source and Background Fields}
\label{subsec:fields}

\ir\ sources are extracted from the \psc\ catalog for source and 
background fields, within a radius of $75\arcsec$ from each 
field center. The background fields are centered on a $3\times 3$ square lattice of
spacing $150\arcsec$ around each source field, and assigned the same
positional uncertainty as the \xray\ source in the source field.  The
square lattice is chosen to ensure that the background fields are
taken from regions similar to the source fields. Background fields
containing a \bsc\ source are excluded from the analysis, since it is
assumed in calculating $R(LR)$ that background fields contain no \ir\
sources that are associated with a \bsc\ source. There are a total of
$150132$ background fields.

\begin{figure}[htb]
\centering
\plotone{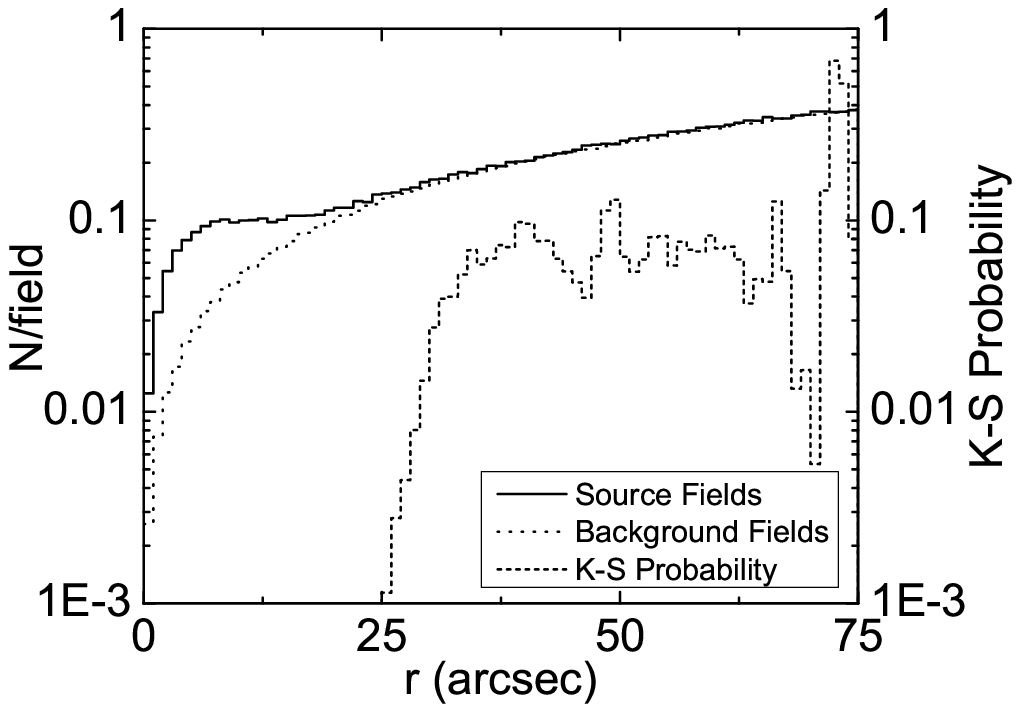}
\caption{[Left Axis] Differential number distributions of \ir\ sources as a function 
of separation $r$ between the \ir\ source and the field center. There is an excess of 
\ir\ sources in the source fields compared to the background fields. 
[Right Axis] At each $r$, the cumulative distributions from $r$ to $75\arcsec$ are 
compared using a two-sample K--S test, which gives the probability that the same 
underlying distribution gave rise to both cumulative distributions. 
The K--S probability is less than $10^{-3}$ for $r<25\arcsec$, demonstrating that 
the excess in source fields is statistically significant (\cf\ \S\ref{subsec:fields}).}
\label{fig:rInt}
\end{figure}

The distance from a \psc\ source to the field center is denoted
$r$. The differential number distributions of separations $r$ 
in the source and background fields are shown in Fig.~\ref{fig:rInt}.
At each $r$, the shapes of the cumulative distributions from $r$ to
$75\arcsec$ are compared using a two sample Kolmogorov--Smirnov
\citep[K--S;][]{press92} test. The K--S test gives the probability that
the cumulative source and background field distributions arise from
the same underlying distribution.  For $r\leq 30\arcsec$ there is an
excess of \ir\ sources in the source fields compared to the background
fields, and the K--S probability is below $10^{-3}$ for
$r<25\arcsec$. This excess in the source fields is due to \psc\
sources that are associated with a \bsc\ source.

There is an improbable difference ($p\sim 4\cdot 10^{-4}$) between the
number distributions of \ir\ sources, observed when comparing source
and background fields in the range $71\arcsec<r<75\arcsec$.  A
detailed examination of these distributions finds a small, but
apparently significant, excess of \ir\ sources in the source
fields compared to background fields.  Only 314 \bsc\
sources have positional uncertainties greater than $30\arcsec$, so it is
unlikely that the excess of \ir\ sources at $r\sim71\arcsec$ is due to 
\psc\ sources associated with a \bsc\ source.  However, the clustering of
\psc\ sources (in open clusters or galaxy clusters) could cause
deviations in the source field distribution compared to that of
background fields, for $r>30\arcsec$, associated with the presence of
the \xray\ source.  
\placefigure{fig:rInt}

\begin{figure}[htb]
\plotone{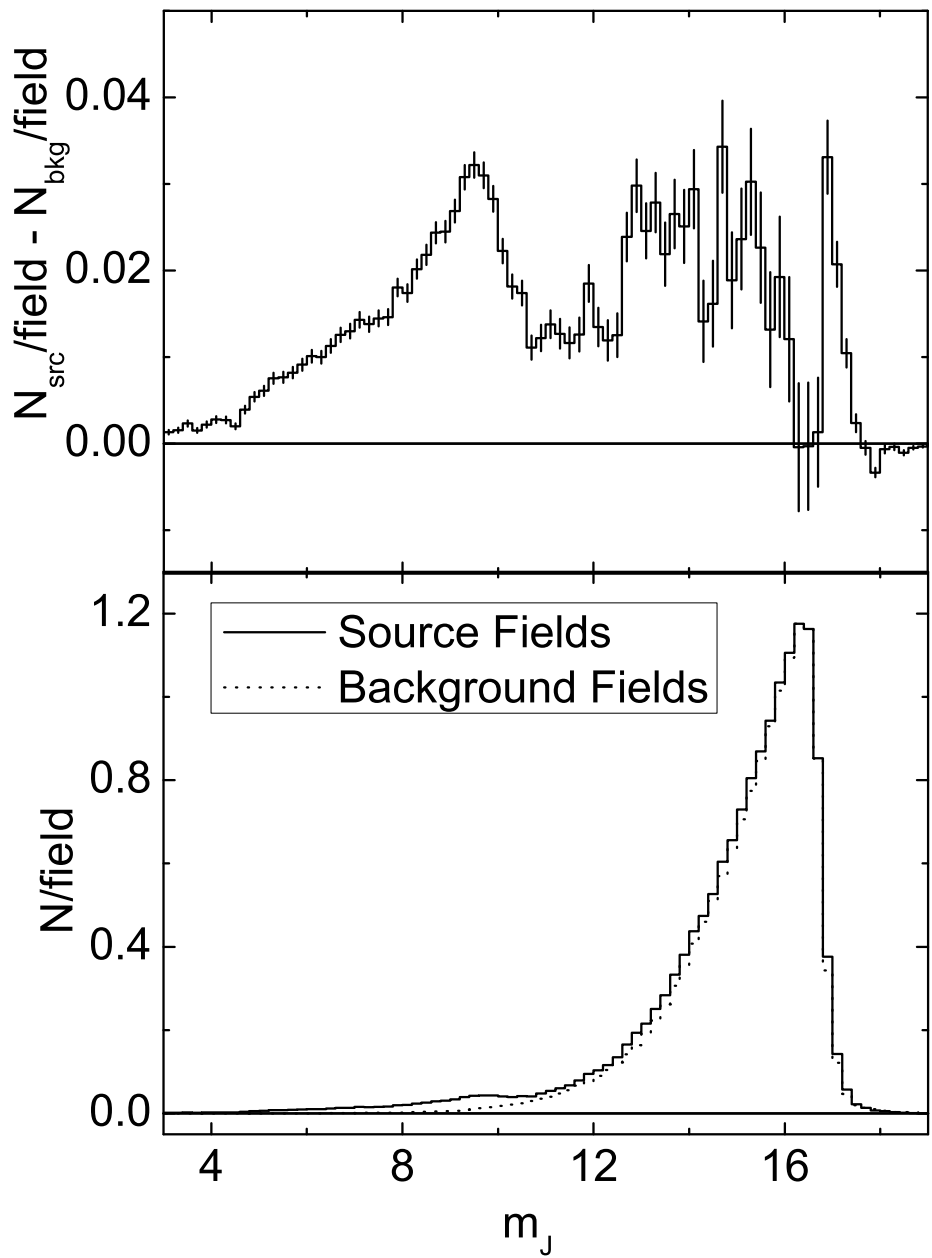}
\caption{[Bottom panel] Differential number distributions in J-band magnitude 
$m_{\rm J}$ of \ir\ sources in the source and background fields. 
[Top panel] The difference between the source and background field distributions. 
There is an excess of \ir\ sources in source fields for $m_J<17.6$, which is 
statistically most significant for $m_J<10.6$.}
\label{fig:jm}
\end{figure}

Differential number distributions in J-band magnitude $m_{\rm J}$ for \ir\ sources in the source and background 
fields are shown in
Fig.~\ref{fig:jm}. There is an excess of \ir\ sources in the source fields, except 
for $m_{\rm J}>17.6$, where there is a slight deficit of \ir\ sources in the source fields compared to the 
background fields. This deficit is likely due to the detectability of faint \ir\ 
sources being limited by the presence of a larger number of bright \ir\ sources in the source fields. 
The excess of \ir\ sources in source fields is statistically most significant for bright ($m_{\rm J}<10.6$) 
\ir\ sources.
\placefigure{fig:jm}

The excess \ir\ sources present in source fields compared to background fields are \ir\ sources 
that are associated in some way with the \xray\ source. 
A large number of known \xray\ source classes are associated with single, point-source 
\ir\ counterparts, like coronally active stars or (spectroscopic or eclipsing) binaries, 
active galactic nuclei (AGN), or low-mass \xray\ binaries. The present analysis is optimized to 
find these sources classes, as they have single \ir\ counterparts. However, there are known \xray\ source 
classes which are associated with multiple \ir\ sources, either due 
to the nature of the source class (\eg\ \xray\ clusters of 
galaxies) or because the source class tends to be found in groups
of similar \ir\ sources (\eg\ T Tauri stars in stellar open 
clusters, or AGN in galaxy clusters). Further, \xray\ emitting \ir\ 
resolved binaries also have more than one \ir\ source 
associated with the \xray\ source. 
These classes of \xray\ sources that are associated with 
multiple \ir\ sources will still appear in the present 
analysis, but the top $P_{\rm id}$ match will be affected by source 
confusion; multiple possible \ir\ counterparts will decrease 
the $P_{\rm id}$ value for the most likely counterpart. 
The remainder of this section attempts to constrain the number of \bsc\ sources that 
are associated with a given number of \ir\ sources. 

\begin{figure}[htb]
\plotone{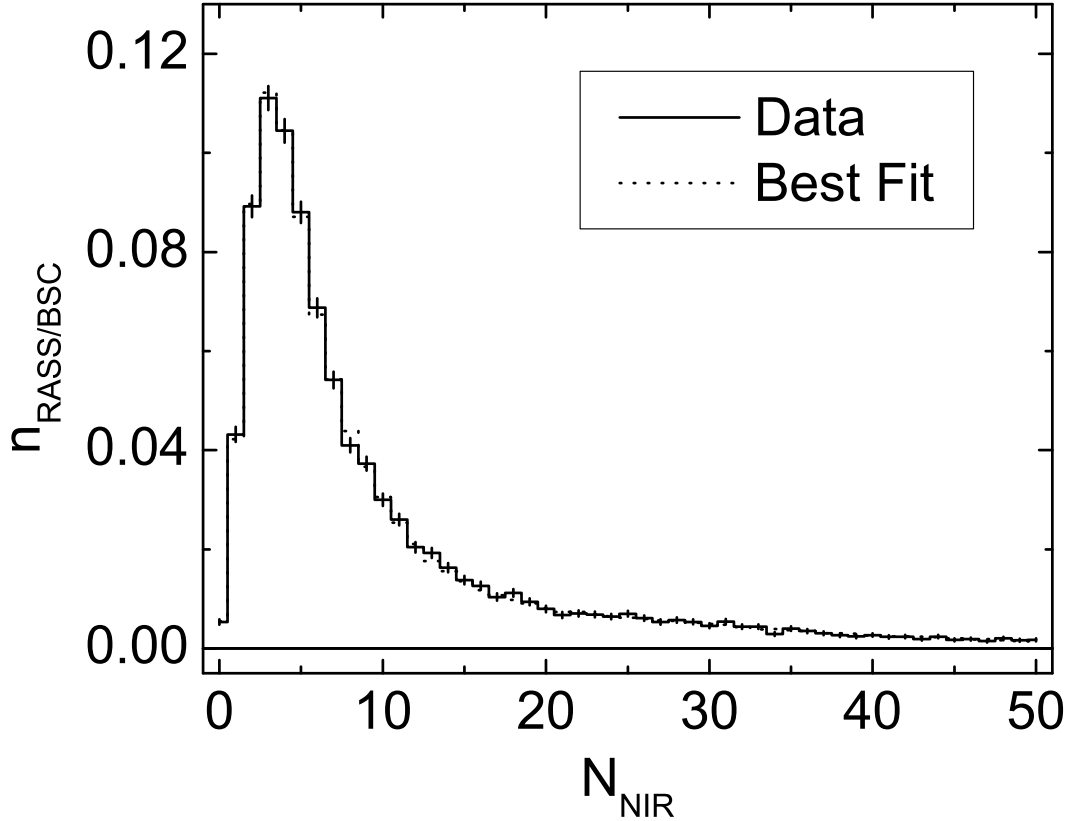}
\caption{Normalized number distribution of \bsc\ sources with $N_{\rm NIR}$ \ir\ sources 
in the corresponding source field, for $N_{\rm NIR}\le 50$, and the best-fit model distribution 
(largely obscured by the data distribution). 
The distribution of \bsc\ sources has a sharp peak around a mode of $N_{\rm NIR}=3$, 
and has a tail extending to $N_{\rm NIR}=373$. There are 1199 \bsc\ sources with $50<N_{\rm NIR}\le 373$, 
with less than 30 \bsc\ sources at any given $N_{\rm NIR}$, which are not included when fitting. 
The best-fit curve (reduced $\chi^2=0.86$) shown 
is for $12\pm1\%$ of \bsc\ sources being associated with 
zero \ir\ sources, $66\pm3\%$ with one, $18\pm3\%$ with two, and $4.0\pm0.6\%$ with seven \ir\ sources 
(\cf\ \S\ref{subsec:fields}).}
\label{fig:nNIR}
\end{figure}

The normalized number distribution of \bsc\ sources with $N_{\rm NIR}$ \ir\ sources 
in the corresponding source field is shown in Fig.~\ref{fig:nNIR}, for $N_{\rm NIR}\le 50$. 
The distribution of \bsc\ sources has a sharp peak around a mode of $N_{\rm NIR}=3$, 
and has a tail extending to $N_{\rm NIR}=373$. There are 1199 \bsc\ sources with $50<N_{\rm NIR}\le 373$, 
with less than 30 \bsc\ sources at any given $N_{\rm NIR}$.
\placefigure{fig:nNIR}

A source field containing a \bsc\ source that is associated 
with $m$ \ir\ sources should on average contain $m$ more \ir\ sources than an adjacent background field. 
Thus, the normalized distribution in $N_{\rm NIR}$ of \bsc\ sources associated with $m$ \ir\ sources 
should be consistent with the normalized number distribution of background fields in $N_{\rm NIR}-m$. 
Further, the overall distribution of \bsc\ sources in $N_{\rm NIR}$ should be consistent with a weighted 
sum over $m$ of background field distributions in $N_{\rm NIR}-m$, where the weights are the 
fraction of \bsc\ sources that are associated with $m$ \ir\ sources. 
To estimate these weights, a Levenberg--Marquardt nonlinear $\chi^2$ minimization fitting \citep[\cf][]{press92} 
to the $N_{\rm NIR}\le 50$ distribution of \bsc\ sources is 
performed using the Origin (OriginLab, Northampton, MA) software. Values of $N_{\rm NIR}>50$ 
are excluded since the uncertainties are assumed to be Gaussian, which is not 
necessarily a good approximation for the low number counts in this range of $N_{\rm NIR}$. 
Since there are almost eight times as many background fields as \bsc\ sources, 
the fractional uncertainties on the background field distribution should in general be less than half 
those on the \bsc\ distribution, though differences in the distributions may increase the 
importance of the background field distribution uncertainties. The background field distribution 
uncertainties are neglected when fitting, and it can be verified after fitting that these uncertainties 
are indeed negligible.

The aim of the fitting is to determine whether there is an acceptable description of the observed 
distribution of \bsc\ sources in $N_{\rm NIR}$, under the assumption that a fraction $f_m$ of all \bsc\ sources 
have $m$ \ir\ counterparts, where $m=1,\, 2,\, 3,\, ...\,$. The method adopted is to first fix $f_m=0$ for all 
$m$ except $m=0$, and then free $f_m$ components one by one (first $m=1$, then $m=2$, etc.) 
until the reduced $\chi^2$ drops below 1. For the $f_m$ that are allowed to vary in each step, a constraint 
$f_m>10^{-10}$ is adopted to avoid computational problems that arise in the fitting routine 
if the constraint $f_m>0$ is used. With the addition of the $m=7$ component, 
the reduced $\chi^2$ value is 0.94, so 
no further components are added. The $m=4,\, 5,\,\&\, 6$ components have best-fit values $f_m=10^{-10}$, \ie\ 
equal to the lower constraint, so these are then fixed to zero. 
Re-running the fit, the $m=3$ component is consistent with zero weight ($f_3=0.02\pm0.04$), 
so this weight is also fixed to zero. The final best fit (reduced $\chi^2=0.86$) is then 
$f_0=0.12\pm0.01$, $f_1=0.66\pm0.03$, $f_2=0.18\pm0.03$, and $f_7=0.040\pm0.006$. 
Thus, $78\pm3\%$ of all \bsc\ sources have zero or one \ir\ counterparts, \ie\ satisfy the assumptions 
used when calculating $P_{\rm id}$ and $P_{\rm no-id}$. Extending the present method to include 
matches with two or more \ir\ sources in the calculation of $P_{\rm id}$ will be the focus of 
future work.

It should be noted that the quoted uncertainties on $f_m$ only 
take into account the diagonal elements of the covariance matrix, and thus do not 
consider dependencies between different $m$ components. Since the different components 
only differ by a shift in $N_{\rm NIR}$ of 1, $f_m$ parameters with similar $m$ are likely to be highly dependent, 
so the true fraction of the \bsc\ that is associated with $m$ \ir\ counterparts may not be consistent with the 
quoted uncertainty on $f_m$. Further, it is likely that the $m=7$ component accounts for a population of 
\bsc\ sources that are associated with $5\la m\la 9$ \ir\ sources, rather than just an $m=7$ population.

Given the best-fit values of $f_m$, it can now be verified that neglecting the uncertainties on the 
background field distribution in $N_{\rm NIR}$ is an appropriate assumption. The largest ratio of the 
uncertainty on the \bsc\ distribution to that on any of the $f_m$ components is 0.27 (for $m=1$ and 
$N_{\rm NIR}=34$). Adding this uncertainty to the \bsc\ distribution uncertainty in quadrature 
would give an uncertainty that is 4\% larger than the \bsc\ distribution uncertainty alone, 
so neglecting the background field distribution uncertainty is unlikely to have a significant effect on the fitting. 

\subsection{$LR$}
\label{subsec:LR}

Large uncertainties in the \bsc\ positions make it difficult to identify likely 
\ir\ counterparts based only on the probability of positional 
coincidence. Therefore, the J-band brightness of the \ir\ source and the local density of 
\ir\ sources were included in the function $LR$: 
\begin{equation}
\label{eq:LR}
LR_i = \frac{\exp(\frac{-r_i^2}{2\sigma^2})\exp(-\rho \pi r_i^2)}{2\pi\sigma^2 N(<m_{{\rm J},i})}\, .
\end{equation}
$N(<m_{{\rm J},i})$ is the number of brighter \ir\ sources in background
fields, $\sigma^2$ is the sum of squares of the uncertainties in the
\xray\ and \ir\ positions, and $\rho$ is the the local density of \ir\
sources.  To compute $\rho$, the number of \psc\ sources within
$30\arcmin$ of the \xray\ source is divided by angular area of
a $30\arcmin$ radius cone.  Some \xray\ sources have a listed positional
uncertainty of zero, for which an uncertainty of 12\arcsec\ is
adopted.

The quantity $LR$ defined in equation~(\ref{eq:LR}) should be understood as
a figure of merit, and not as a likelihood ratio in the formal
statistical definition of that term \citep[\cf][]{wilks38}; 
in the formal definition the likelihood ratio is a ratio of probability
density functions, whereas the quantity $LR$ as used here and
previously \xidp\ includes integrated probability density functions,
and thus does not satisfy this definition. Moreover, the quantity
$LR$ is not here used as a likelihood ratio, in that a likelihood
ratio is used to determine the relative likelihood of two competing
hypotheses; here, $LR$ is used strictly as a figure of merit and not,
itself, to compare the relative likelihood of two competing
hypotheses.\footnote{The quantity $LR$ evolved \xidp\ as a modified
version of a quantity referred to as a ``likelihood ratio'', used in
previous work \citep{wolstencroft86} to identify likely off-band
counterparts of IRAS sources.  The terminology ``likelihood ratio''
was applied in the previous work \citep{wolstencroft86}, where
the relevant quantity was in fact used to choose between two competing
hypotheses; however, the quantity to which it was applied involved
terms which were not probability density functions, so the quantity used there does
not satisfy the formal definition of a likelihood ratio.  Thus, the reader
is strongly cautioned that in the present work, the quantity $LR$ is used strictly as
a figure of merit; it is not itself used as a likelihood ratio to
distinguish between two competing hypotheses, and should not be 
considered as anything other than strictly a figure of merit.}

The $LR$ used in the present work differs from that employed in the
\bsc--\usno\ cross-association by a factor of $(2\pi\sigma)^{-1}$, due
to the use of a two dimensional---rather than one
dimensional---normalization of the probability of coincidence, and by
the added factor of $\exp(-\rho \pi r_i^2)$, which is the probability
that there is no closer \ir\ source to the field center. $\rho$ ranges
from 0.3 to 50 \ir\ sources per square arcmin, is 2.8 sources per
square arcmin on average, and is typically large close to the galactic
plane and small away from it. Including a $\rho$ dependence in $LR$
accounts for the fact that the higher the local density of \ir\
sources, the larger the chance that an unrelated background \ir\
source falls within the error cone of an \xray\ source.

\begin{figure}[htb]
\plotone{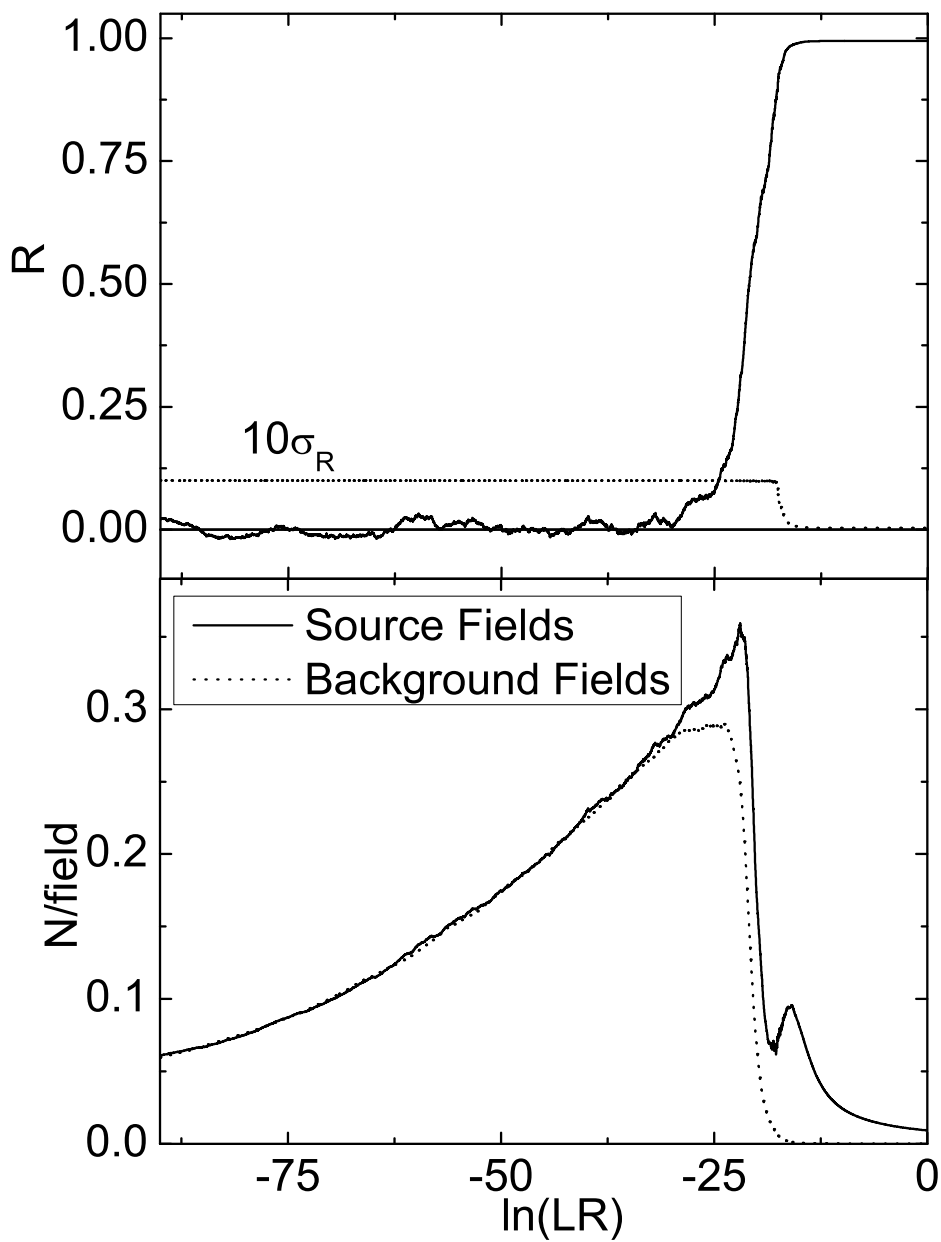}
\caption{[Bottom panel] Number distributions in $LR$ of \ir\ sources in 
source and background fields. 
The normalization of the number distributions is such that 
$\int (N/{\rm field}) d(\ln(LR))$ approximately equals the total number of 
\ir\ sources per field for that field type (the equality is not exact because of the 
averaging over bins described in \S\ref{subsec:LR}). 
[Top panel] The calculated $R(LR)$. For $\ln(LR)<-34.07$, $R(LR)$ is set to zero.}
\label{fig:R}
\end{figure}

Fig.~\ref{fig:R} shows the number distributions in $LR$ of \ir\ sources in
source and background fields, and the resulting $R(LR)$.  To ensure sufficient
counting statistics, the number distributions are calculated by first
dividing the $\ln(LR)$ range into bins of width 0.01, and then for each
bin averaging over a wide enough range of bins around it---at least
21---to keep the propagated Gaussian counting uncertainty $\sigma_R<0.01$, and to keep
the total number of \ir\ sources above 250---to justify using the
Gaussian uncertainty. 
The normalization of the number distributions is such that 
$\int (N/{\rm field}) d(\ln(LR))$ approximately equals the total number of 
\ir\ sources per field for that field type (the equality is not exact because of the 
averaging over bins described above). 
$R(LR)$ is 
set to zero for $LR$ values less than the largest $LR$ for which $R(LR)=0$, 
so that $R\geq0$ for all \xray--\ir\ pairs. Negative
$R$ values can arise from random fluctuations, or from breakdown of the
assumption that source fields contain both background
\ir\ sources and \ir\ sources that are associated with the \xray\
source. An example of such breakdown is the apparent deficit of very
faint \ir\ sources in source fields compared to background fields.
\placefigure{fig:R}

\xray--\ir\ pairs with $ln(LR_i)$ values in the same bin are all assigned 
the same $R_i$ value, so $R_i$ values in the same $ln(LR_i)$ bin are perfectly 
correlated. Further, the averaging over 
bins used to calculate $R(LR)$ introduces a correlation between bins. 
All bins with $ln(LR)\ge -16.05$ are correlated to some extent, since there
are only 247 \ir\ sources in background fields with $LR$ in this range, 
and any bin in that range of $LR$ thus must include at least one bin 
with $ln(LR)<-16.05$ in the averaging described above. For the remaining range 
of $LR$ where $R(LR)$ is non-zero, $-34.07<ln(LR)<-16.05$, the distance in 
$ln(LR)$ over which a correlation exists varies between 0.23 and 2.33. 
There are 10191 source fields with at least one pair of matches for which the 
$R$ values are not independent. Thus, care must be taken when propagating 
$\sigma_{\rm R}$ to calculate $\sigma_{\rm Pid}$, and if the uncertainties are propagated 
as if the $R$ values are independent, it must be verified that this assumption 
does not significantly affect the $\sigma_{\rm Pid}$ values of interest.

The chosen form of $LR$ neglects any dependence on $\rho$ of the number
distribution of \ir\ sources as a function of $m_{\rm J}$. Such a dependence
could arise from source population differences between the denser
galactic plane and the less dense regions at higher galactic
latitudes, or from the detectability of faint sources being limited by
the presence of bright sources.  Some dependence on $\rho$ is seen in 
a normalized version of $N(<m_{{\rm J,}i})$, and this is most significant for
the very brightest sources ($m_{\rm J}<4$). However, there are very few
$m_{\rm J}<4$ sources in source or background fields (\cf\ 
Fig.~\ref{fig:jm}), and these are likely to have $R$ values close to 1 
when they appear in source fields (\cf\ \S\ref{sec:results}), so it is
not expected that addressing these population differences would have a
significant impact on the resulting association catalog.

\subsection{Control Fields}
\label{subsec:control}

In addition to the source and background fields, one control field
($75\arcsec$ radius) is extracted per square degree of the celestial
sphere. Control fields that overlap with a source field are excluded,
bringing the total number of control fields to 41192.  Each control
field is assigned the properties of a randomly chosen \bsc\ source,
and $LR_i$, $R(LR_i)$, and $P_{{\rm id,}i}$ are calculated for each \ir\
source in a given control field using the $R(LR)$ function computed
from the source and background fields.  Note that $R(LR_i)$ is a
function which depends only on the source and background fields, not on 
the control fields.

A control field resembles an \xray\ source that is not associated with a
\psc\ source, so the $P_{\rm no-id}$ values of control fields should
be comparable to those of near-infrared faint \bsc\ sources, and to
\bsc\ sources associated with any extended \ir\ sources that do not
appear in the \psc.  This information has proven useful in population
studies of blank-field X-ray sources, such as isolated neutron stars
\citep{rutledge03, turner09}.

\section{RESULTS}
\label{sec:results}

For 100 \bsc\ sources, there are no \psc\ sources within $75\arcsec$. For the remaining 18706 
\xray\ sources, there are a total of 287736 distinct \ir\ sources within $75\arcsec$. 
Except for three \ir\ sources that lie within $75\arcsec$ of both 1RXS~J020220.5-010609 and 
1RXS~J020221.0-01071, each \ir\ source is paired with only one \bsc\ source, so 
the distinction between an \ir\ source and an \xray--\ir\ pair can be relaxed. 
Assuming Gaussian counting uncertainty on the number of \ir\ sources, there 
are $15.30\pm0.03$ \ir\ sources per source field, and $14.18\pm0.01$ per background field, 
amounting to an excess in source fields of $1.12\pm0.03$ \ir\ sources per field.

The distribution in $LR$ of the excess of \ir\ sources in source fields can be seen from 
Fig.~\ref{fig:R}. For $\ln(LR)<-34.07$, the source and background field distributions are consistent with 
no excess in source fields ($0.03\pm0.03$ sources per field), 
and the probability that the same underlying distribution would give rise to 
two distributions as or more different than the source and background field distributions---for 
$\ln(LR)<-34.07$---is 0.55 \citep[two sample K--S test;][]{press92}. Thus, the number distributions of \ir\ sources in 
source and background fields are consistent with being identical for $\ln(LR)<-34.07$, and 
setting $R(LR)=0$ for this range of $LR$ is justified. 
For $\ln(LR)\ge-34.07$, the excess of \ir\ sources in source fields compared to background fields 
is $1.10\pm0.02$ per field. Thus, there are $20600\pm300$ \ir\ sources in source fields 
in excess of the number of expected background sources for $\ln(LR)\ge-34.07$, and these 
\ir\ sources are assumed to be associated with the \bsc\ sources when calculating $R(LR)$. 

Background fields are positioned near source fields to ensure that the population 
of \ir\ sources in background fields is similar to the population of background 
\ir\ sources in source fields, since the two populations are assumed to be identical 
when calculating $R(LR)$. Further, the positional uncertainty of the \bsc\ source in a 
given source field is adopted for the adjacent background fields, so that any anisotropy 
in the distribution of \bsc\ positional uncertainties also is present for the background fields. 
The demonstrated similarity of the source and background field distributions of \ir\ sources 
with $\ln(LR)<-34.07$ indicates that the background field \ir\ population 
does indeed approximate the background population in source fields, most likely also for 
$\ln(LR)\ge-34.07$. 

There are $15.62\pm0.02$ \ir\ sources per control field, so there are more 
\ir\ sources per control field than per source field or per background field.
Table~\ref{table:skypos} gives the number of source and control fields in six different 
regions of the sky, and the average value of $\rho$---the local density of NIR sources---for the fields in that region. 
The regions are chosen to differentiate between the galactic poles, center, 
and anti-center, since these have different average \ir\ source
densities in the \psc\ catalog. 
A larger fraction of control fields than source fields lie in the galactic plane, where 
the \ir\ source density is larger than in the remaining part of the sky, 
and this explains why there are more \ir\ sources per control field than per source field or per background field. 
Only \ir\ sources with $\ln(LR)\ge-34.07$ are of interest in the case of control fields, 
since only these sources have $R>0$, and for this range of $LR$ 
the cumulative distributions of \ir\ sources in control and background fields 
are consistent with arising from the same underlying distribution 
\citep[two sample K--S test probability of 0.19;][]{press92}. The distributions thus only 
differ in normalization for this range of $LR$.
\placetable{table:skypos}

\begin{figure}[htb]
\plotone{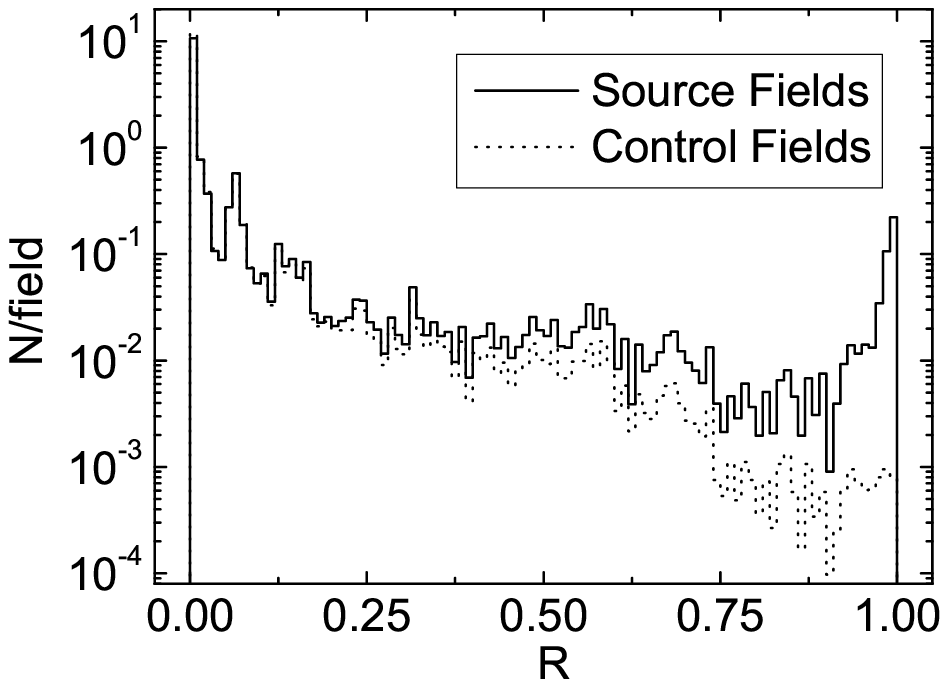}
\caption{Differential number distributions in $R$ of \ir\ sources in the source
and control fields. 
For $R\ge0.10$ the number density in source fields is above that in control fields, 
while for $R<0.10$ the number density in source fields is below that in control fields in all but two bins. 
In total, there is a 9\% excess of \ir\ sources in control fields compared to source fields for $R<0.10$, 
and a 78\% excess in source fields compared to control fields for $R\ge 0.10$ (\cf\ \S\ref{sec:results}).
}
\label{fig:L}
\end{figure}

\begin{figure}[htb]
\plotone{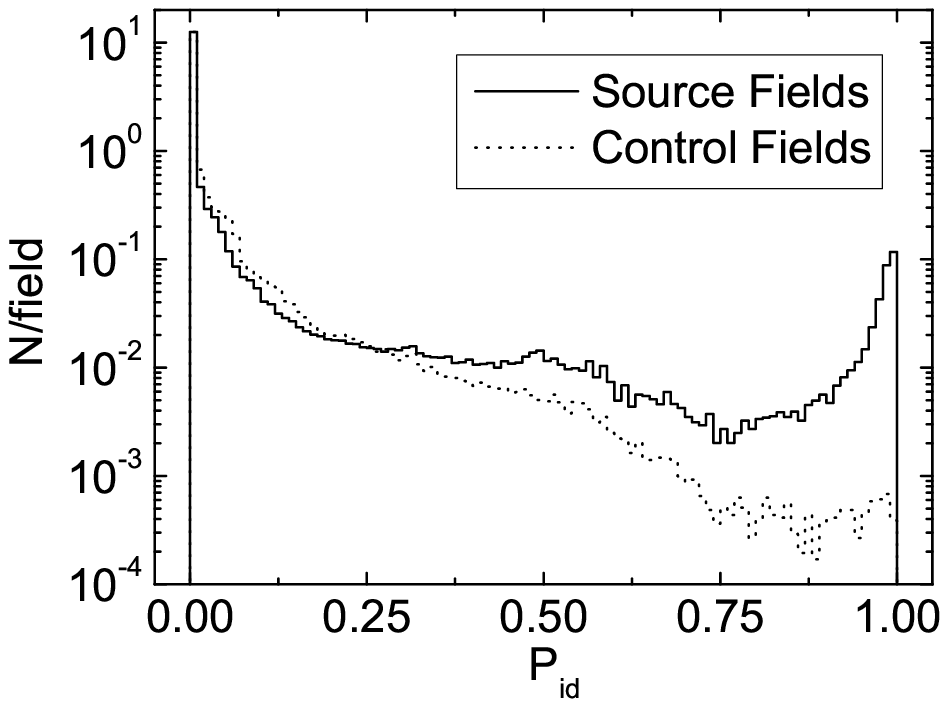}
\caption{Differential number distributions in $P_{\rm id}$ for \ir\ sources in  the source and control fields. 
For $P_{\rm id}\ge 0.28$ the number density in source fields is above that in control fields, 
while for $P_{\rm id}<0.28$ the number density in source fields is below that in control fields in all but two bins. 
In total, there is a 6\% excess of \ir\ sources in control fields compared to source fields for 
$P_{\rm id}<0.28$, and a 222\% excess in source fields compared to control fields for $P_{\rm id}\ge 0.28$ 
(\cf\ \S\ref{sec:results}).
}
\label{fig:Pid}
\end{figure}

To examine the impact of using $P_{\rm id}$ as a measure of the
significance of an association rather than $R$, it is useful to
consider the differential number distributions in $R$ and $P_{\rm id}$
of \ir\ sources in the source and control fields. The distributions in
$R$ are shown in Fig.~\ref{fig:L}, while the distributions in $P_{\rm
id}$ are shown in Fig.~\ref{fig:Pid}. 
For $R\ge0.10$ the number density in source fields is above that in control fields, 
while for $R<0.10$ the number density in source fields is below that in control fields in all but two bins. 
In total, there is a 9\% excess of \ir\ sources in control fields compared to source fields for $R<0.10$, 
which is due to the larger average number of \ir\ sources in control fields. For $R\ge 0.10$, there 
is a 78\% excess of \ir\ sources in source fields compared to control fields, and this excess is due to 
\ir\ sources that are associated with a \bsc\ source only being present in source fields.
The distributions in $P_{\rm id}$ show that for $P_{\rm id}\ge 0.28$ 
the number density in source fields is above that in control fields, 
while for $P_{\rm id}<0.28$ the number density in source fields is below that in control fields in all but two bins. 
In total, there is a 6\% excess of \ir\ sources in control fields compared to source fields for 
$P_{\rm id}<0.28$, and a 222\% excess in source fields compared to control fields for $P_{\rm id}\ge 0.28$.
If each field contained only one $R>0$ source, $P_{\rm id}$ would
equal $R$ for all \ir\ sources, and the distributions in $P_{\rm id}$ would be identical to the 
distributions in $R$ (\cf\ eq.~[\ref{eq:P.id.i.expl}]). However, 85\% of source fields and
58\% of control fields contain more than one $R>0$ source, so the distributions do 
in fact differ. 
The reason that a larger fraction of source fields than control fields contain 
multiple \ir\ sources with $R>0$ is because source fields contain \ir\ sources that 
are associated with a \bsc\ source, while control fields do not. 
As a result, unrelated background sources 
in source fields are assigned lower $P_{\rm id}$ values than similar sources in control fields, 
and it is possible that $n_{\rm control}(P_{\rm id})>n_{\rm source}(P_{\rm id})$ for some $P_{\rm id}=R$ where 
$n_{\rm control}(R)\leq n_{\rm source}(R)$, as is the case for $0.10\leq P_{\rm id}<0.28$. 
\placefigure{fig:L}
\placefigure{fig:Pid}

The quality of the association catalog (eq.~[\ref{eq:Q}]) is $Q=0.70\pm0.02$. 
The uncertainty on $Q$ is a conservative estimate of the statistical uncertainty, 
computed by noting that $Q$ is simply the sum of 
of all $P_{\rm id}$ values in source fields divided by the number of \bsc\ sources. 
Since each $P_{\rm id}$ value is correlated with at least 249 other source field 
$P_{\rm id}$ values (due to the binning described in \S\ref{subsec:LR}), their 
$\sigma_{\rm Pid}$ values cannot simply be summed in quadrature as if the 
$P_{\rm id}$ values are independent. Rather, the conservative assumption that 
all $P_{\rm id}$ values are perfectly correlated is adopted, and the uncertainties summed absolutely 
to obtain the conservative estimate of the statistical uncertainty on $Q$.

The interpretation of $Q$ is that it is the fraction of the \bsc\ with a counterpart 
in the \psc\ identifiable by the present method \xidp. Therefore, 
$1-Q$ is the fraction of the \bsc\ that does not have an \ir\ counterpart, 
under the assumption that all \bsc\ sources have zero or one \ir\ counterparts. 
However, it was shown in \S\ref{subsec:fields} that this assumption only holds 
for $78\pm3\%$ of \bsc\ sources, so the assumption may introduce a large systematic 
uncertainty on $1-Q$. In fact, the fraction of \bsc\ sources found in \S\ref{subsec:fields} to have no 
\ir\ counterpart is $12\pm1\%$, which is not consistent with the statistical uncertainty on $1-Q$. 
Thus $Q$, and by extention many $P_{\rm no-id}$ values, have large systematic uncertainties introduced 
by the assumption of zero or one \ir\ counterparts to each \bsc\ source. 
These systematic uncertainties are also likely to be present in the 
previous work \xidp, as it was concluded there that $\sim27.2\%$ of 
\bsc\ sources do not satisfy the assumption that each \xray\ source has 
zero or one \usno\ counterparts.

\begin{figure}[htb]
\plotone{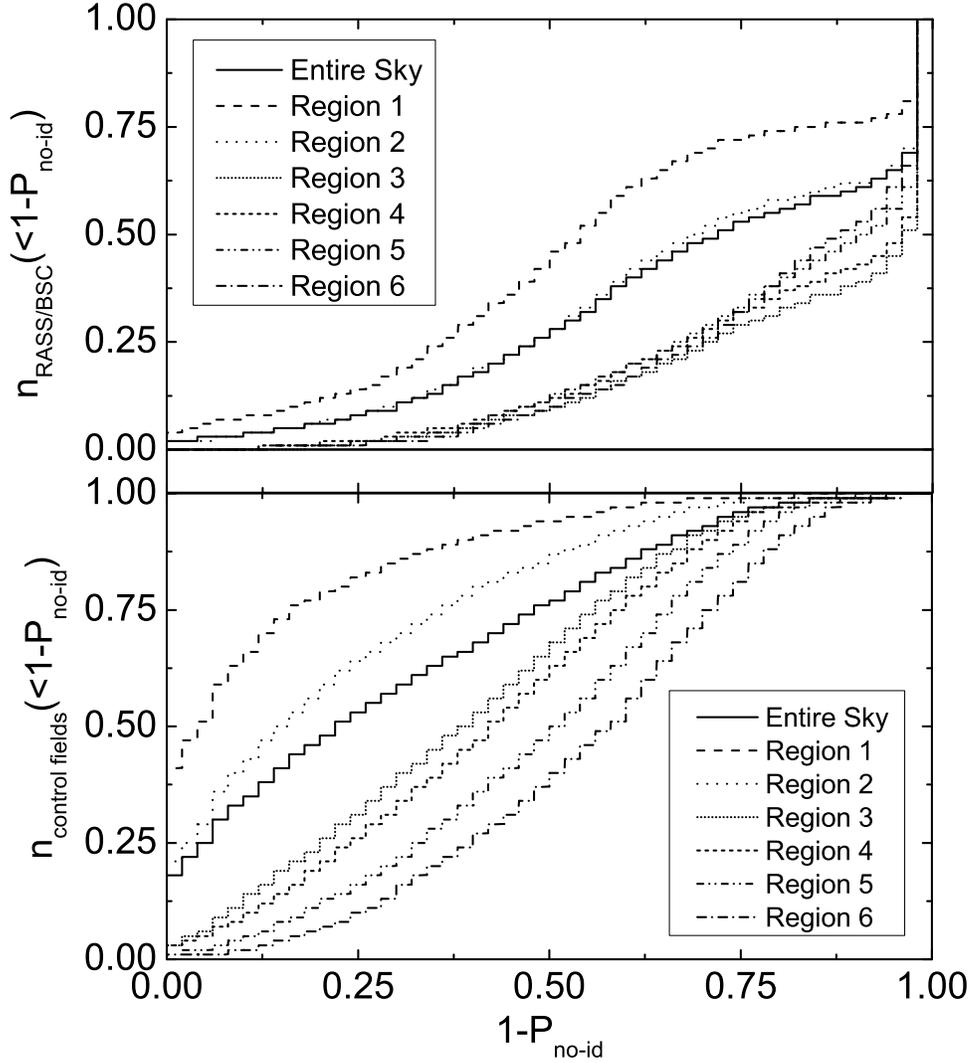}
\caption{[Bottom panel] Normalized cumulative distributions in $1-P_{\rm no-id}$ of control fields 
in six different regions of the sky (\cf\ Table \ref{table:skypos}). 
See \S\ref{sec:results} for further discussion. 
[Top panel] Normalized cumulative distributions in $1-P_{\rm no-id}$ of \bsc\ sources in 
the same six regions as for the control fields. The distributions for the regions in the galactic 
plane follow each other closely for $1-P_{\rm no-id}<0.75$, and lie entirely below the distributions 
for regions outside the galactic plane.}
\label{fig:pnoid}
\end{figure}

Control fields do not contain \ir\ sources that are associated with a 
\bsc\ source, but have $P_{\rm id}$ and $P_{\rm no-id}$ values computed 
based on the $R(LR)$ function computed using the source and background 
fields. The control fields are thus like the source fields of 
\xray\ bright \ir\ faint sources in the \bsc. 
Fig.~\ref{fig:pnoid} shows the
normalized cumulative distributions in $1-P_{\rm no-id}$ of
\bsc\ sources and control fields, $n(<1-P_{\rm no-id})$, for the six
different regions of the sky described in Table~\ref{table:skypos}.
At every value of $1-P_{\rm no-id}$, $n_{\rm control\ fields}(<1-P_{\rm
no-id})$ is larger for each region than for any other region with
larger average \ir\ source density. Thus, \xray\ bright \ir\ faint
sources are more easily identified in regions of lower \ir\ source
density.  In the case of \bsc\ sources, the distributions for regions
in the galactic plane follow each other closely for $1-P_{\rm
no-id}<0.75$, and lie entirely below the distributions for regions
outside the galactic plane. This may in part be due to greater
confusion in the galactic plane as a result of higher \ir\ source
density, but is most likely indicative of a difference in the
population of \xray\ sources in and outside the galactic plane.
\placefigure{fig:pnoid}

\subsection{The Association Catalog}
\label{subsec:catalog}

Of the 18806 \bsc\ sources, 18568 are associated with at least one \psc\ source 
with $P_{\rm id}>0$. Further, each \bsc\ source is assigned a $P_{\rm no-id}$. 
The association catalog (Table \ref{table:catalog}) contains all 
\bsc\ sources, their $P_{\rm no-id}$, and the \psc\ source in the corresponding source 
field with the highest non-zero $P_{\rm id}$. If any of the \bsc\ sources or \psc\ sources 
in the association catalog 
were listed in the \simbad\ database as of 2009 April 23, the source type given there is also included 
(\cf\ \S\ref{sec:prevclass} for details). The association catalog is sorted by the 1RXS designation 
of the \bsc\ source.
An online version of the catalog is also 
available\footnote{\url{http://dualcore.physics.mcgill.ca/RASS/index.html}}, and is 
searchable by $P_{\rm id}$ and $P_{\rm no-id}$.
\placetable{table:catalog}

The catalog contains 3853 \twomass\ sources with $P_{\rm id}>0.98$, 2280 with
$0.98\ge P_{\rm id}>0.9$, and 4153 with $0.9\ge P_{\rm id}>0.5$. 
These subsets of the catalog are referred to as high, medium, and low quality matches, 
respectively, and correspond to the catalogs which resulted from the \bsc--\usno\ cross-association 
\xidp. The background contamination ($N_{\rm bkg}=N-\Sigma_iP_{{\rm id,}i}$, \cf\ \xidt) 
is 39, 101, and 1443 for the high, medium, and low quality matches respectively.

\begin{figure}[htb]
\plotone{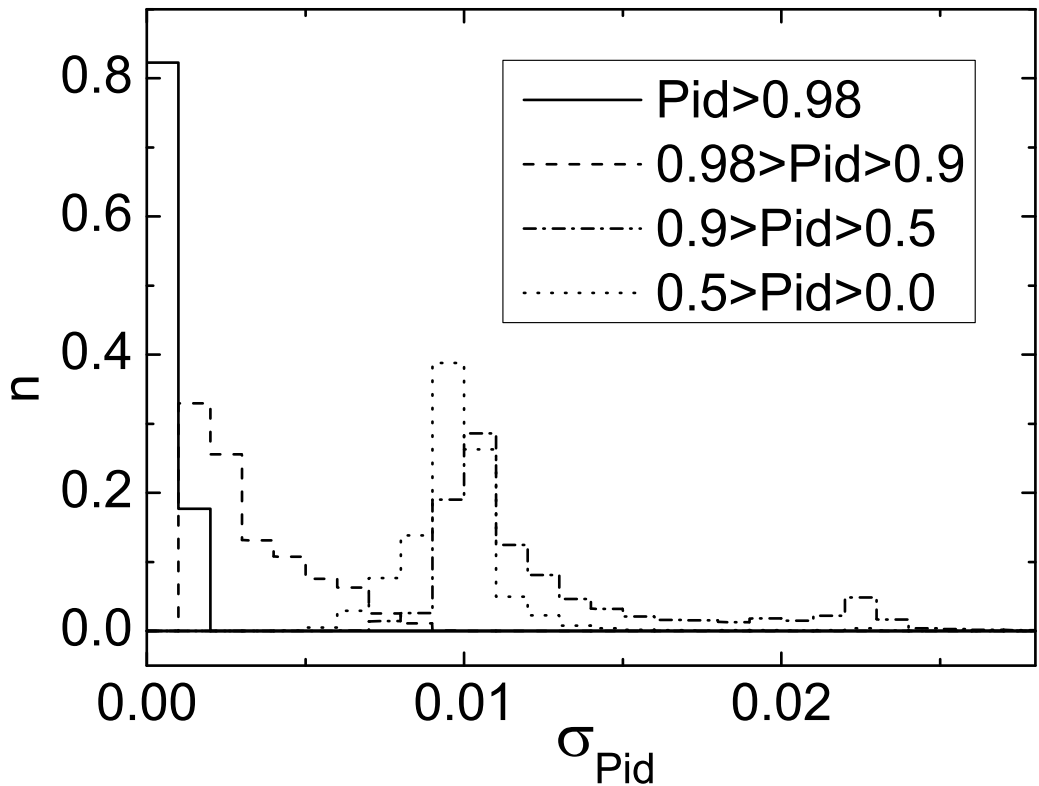}
\caption{Normalized number distributions in $\sigma_{\rm Pid}$ of \xray--\ir\ matches in the association 
catalog. The high and medium quality matches all have $\sigma_{\rm Pid}<0.01$, while the low quality 
matches have $\sigma_{\rm Pid}$ values clustered around 0.01, with a tail in the distribution extending 
to 0.03. For $P_{\rm id}\le0.5$ almost all matches have $\sigma_{\rm Pid}$ values 
clustered around 0.01. See \S\ref{subsec:catalog} 
for further discussion.}
\label{fig:dPid}
\end{figure}

The distributions of statistical uncertainties $\sigma_{\rm Pid}$ on $P_{\rm id}$ are shown in Fig.~\ref{fig:dPid}. 
The high and medium quality matches all have $\sigma_{\rm Pid}<0.01$, while the low quality 
matches have $\sigma_{\rm Pid}$ values clustered around 0.01, with a tail in the distribution extending 
to 0.03. For $P_{\rm id}\le0.5$ almost all matches have $\sigma_{\rm Pid}$ values clustered around 0.01.
The clustering around $\sigma_{\rm Pid}\approx0.01$ is due to the bin size used in calculating $R(LR)$ 
being adjusted until $\sigma_{\rm R}<0.01$, since the uncertainty in $R$ is approximately the 
uncertainty in $P_{\rm id}$ for sources where $R\approx P_{\rm id}$. The tail of larger $\sigma_{\rm Pid}$ 
consists of sources for which there is a significant difference between $R$ and $P_{\rm id}$ due to the 
presence of other sources with non-zero $R$ values in the same field.
\placefigure{fig:dPid}

In propagating the uncertainty $\sigma_{\rm R}$ to $\sigma_{\rm Pid}$, it is assumed that 
the $R$ values in that field are independent. However, due to the method used to calculate $R(LR)$, 
that may not be the case (\cf\ \S\ref{subsec:fields}). Thus, it must be verified that treating 
the $R$ values as independent has not led to significantly smaller $\sigma_{\rm Pid}$ values 
than those that would result from an error analysis considering correlations. An investigation 
of the matches in the association catalog reveals that none of the high or medium quality matches 
have $R$ values that are correlated with another $R$ value in the field, while 445 low quality matches 
and 2799 $P_{\rm id}\le0.5$ matches do have $R$ values that to some extent are correlated with at least 
one other $R$ value in the field. 
All of these low quality matches have $R$ values that are correlated with exactly one other $R$ value in 
the field, while 262 of the $P_{\rm id}\le0.5$ matches have $R$ values that are correlated with 
between two and five other $R$ values. A conservative rough estimate of the effect of correlations in the 
error propagation is that $\sigma_{\rm Pid}$ is a weighted absolute sum rather than the square root of 
a weighted sum of squares, so considering correlations could increase $\sigma_{\rm Pid}$ by as much as 
a factor of square root the number of correlated $R$ values. 
Thus, the uncertainties shown for the high and medium quality matches 
in Fig.~\ref{fig:dPid} should be largely unaffected by the assumption that the $R$ values are independent, 
while 11\% of the low quality matches may have quoted uncertainties smaller by as much as a factor of $1/\sqrt{2}$ 
than those that would result from a covariant error analysis. Further, 31\% of the $P_{\rm id}\le0.5$ matches 
may have uncertainties that are underestimated by as much as a factor of $1/\sqrt{2}$, 
while 3\% may be underestimated by as much 
as a factor of $1/\sqrt{6}$. Based on the distributions shown in Fig.~\ref{fig:dPid}, almost all 
low quality and $P_{\rm id}\le0.5$ matches would thus have $\sigma_{\rm Pid}<0.03$, also in a 
covariant error analysis. 
Whether the uncertainty on a low quality or $P_{\rm id}\le0.5$ match is $0.01$ or $0.02$ is unlikely to affect 
a decision to do follow-up observations of the \ir\ or \xray\ source, nor is it likely to affect any 
conclusions drawn on the basis of the association, 
so it is concluded that the assumption of independent $R$ values is adequate for the present work.

In principle, correlations between two other $R$ values in the field of a high or medium quality match could 
lead to a larger $\sigma_{\rm Pid}$ value in an error analysis considering correlations. 
However, those $R$ values are likely to be small compared to the $R$ value of the association catalog match 
(otherwise it would not be a high or medium quality match), so the contribution of the uncertainties on 
the other $R$ values to $\sigma_{\rm Pid}$ is likely to be small, whether or not the other $R$ values 
are correlated. Thus, the conclusion remains that the uncertainties shown for the high and medium quality matches 
in Fig.~\ref{fig:dPid} should be largely unaffected by the assumption that the $R$ values are independent. 
However, the uncertainties for the low quality and $P_{\rm id}\le0.5$ matches may be affected by correlations 
between other $R$ values in the field, so the fractions of these matches that are affected may be larger 
than those quoted above, as may the size of the effect in some cases. 
Fields for which there are many correlated $R$ values are likely to be fields for which 
the assumption of one or no \ir\ counterparts to the \bsc\ source is not valid, 
so the systematic uncertainty introduced by this assumption is likely to be larger than 
the propagated statistical uncertainty for these fields. Thus, whether the uncertainty is 
propagated as if the variables are independent or not would not significantly affect the 
quantitative or qualitative conclusions under the present analysis.

\begin{figure}[htb]
\plotone{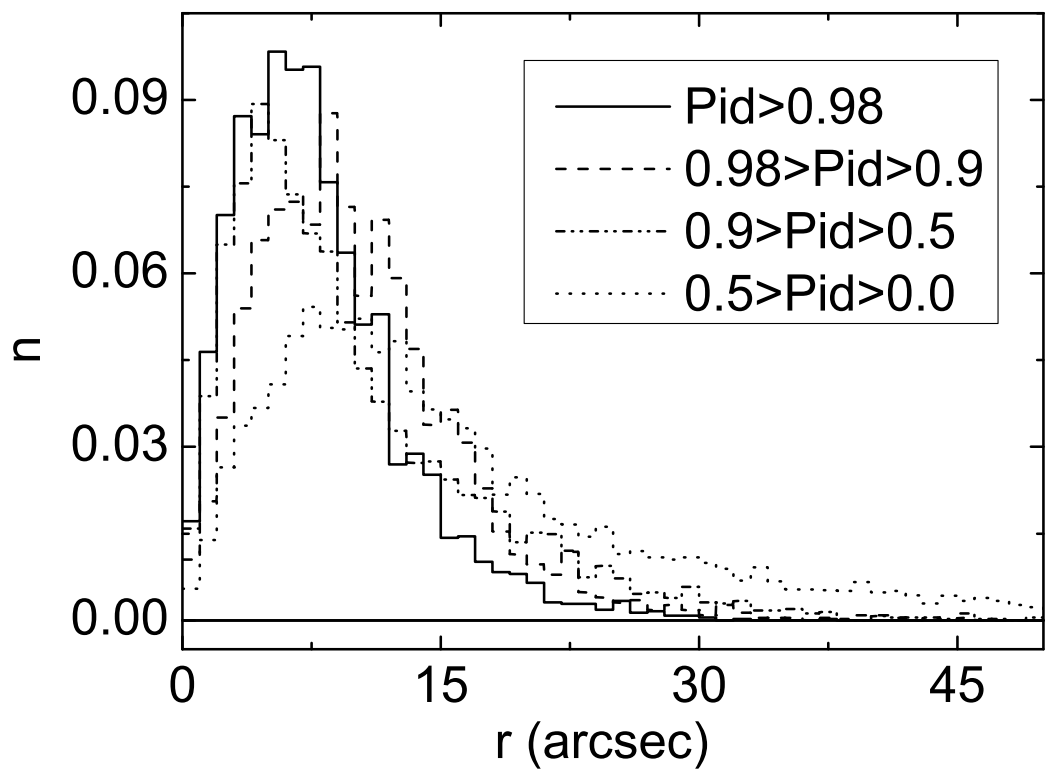}
\caption{Normalized number distributions in separations $r$ of the \xray--\ir\ matches in the association 
catalog. The distribution for high quality matches is the most sharply peaked (with a mode at $6\arcsec$), 
and a broadening is seen with decreasing $P_{\rm id}$. 
See \S\ref{subsec:catalog} for discussion.}
\label{fig:rPid}
\end{figure}

The distribution of separations $r$ between the \xray\ and \ir\ sources in the association catalog is shown in 
Fig.~\ref{fig:rPid}. The distribution for high quality matches is the most sharply peaked, 
and a broadening is seen as $P_{\rm id}$ decreases. High, medium, and low quality matches are 
identified out to a distance comparable to the positional uncertainty for the \bsc\ 
sources; the mode and average \bsc\ positional uncertainty are $10\arcsec$ and $12.7\arcsec$ 
respectively. 
Several $P_{\rm id}<0.5$ matches are identified well beyond this range, some at distances 
greater than $45\arcsec$.
\placefigure{fig:rPid}

Fig.~\ref{fig:jmPid} shows the distribution of J-band magnitudes among the matches in the catalog. 
Almost all high and medium quality matches have $m_{\rm J}<12$, while the low quality and remaining matches 
include a population of significantly fainter sources. As will be discussed in \S\ref{subsec:color}, 
many of the $m_{\rm J}<12$ sources are coronally active stars, and the sharp drop in the number of 
matches at $m_{\rm J}\sim 10$ is likely due to coronally active stars fainter than this $m_{\rm J}$ having 
an \xray\ flux below the \bsc\ detection limit. The second, J-band fainter, population of matches seen 
in Fig.~\ref{fig:jmPid} appears to match the brightness distribution of \ir\ sources in the background fields, 
indicating that there is no strong \xray\ to J-band flux correlation for these matches.
\placefigure{fig:jmPid}

\section{SOURCE CLASSES IN \simbad}
\label{sec:prevclass}

The \simbad\ database\footnote{\url{http://simbad.u-strasbg.fr/simbad/}} lists source types 
for many astronomical objects, including---as of 2009 April 23---all 18806 \bsc\ sources, and 
6115 of the \psc\ sources in the association catalog (Table~\ref{table:catalog}). 
\simbad\ is an incomplete, heterogeneous database, but offers a sense of the extent to which the association 
catalog includes previously unknown information. 
Objects listed in \simbad\ with the source type `X-ray source' or 'Infra-Red source' are considered 
unclassified, and these include 12213 (65\%) of the \bsc\ sources and 229 of the 
aforementioned \psc\ sources. The remaining 6553 \bsc\ sources and 5886 \psc\ sources 
are considered classified in \simbad. 
Of the 18568 matches in the association catalog, 2609 appear as associated in \simbad\ 
(the \bsc\ and \psc\ identifiers are both listed for the same object). 
This section examines the source types and \bsc--\psc\ associations listed 
in \simbad\ for the \psc\ sources in the association catalog.

All \bsc--\psc\ matches fall into one of four mutually exclusive categories: 
a \bsc\ source and a \psc\ source may or may not be associated in \simbad, and 
the \psc\ source may or may not be classified in \simbad. 
The number of matches in the association catalog that fall into each of these 
categories is given in Table~\ref{table:SIMBAD}, for several different $P_{\rm id}$ cutoffs.
For the classified \psc\ sources that are not associated with the 
corresponding \bsc\ source in \simbad, the \psc\ source type can 
be adopted for the \bsc\ source with confidence $P_{\rm id}$. 
Moreover, this table points out where opportunities to classify \bsc\ sources exist, 
through observations of the \psc\ source. For example, there are 1203 high 
quality matches for which the \psc\ source is not classified in \simbad, and follow-up 
optical/IR spectroscopy of these \psc\ sources could classify the \xray\ source 
as an AGN, star, white dwarf, or other class of \xray\ source.
\placetable{table:SIMBAD}

The number of \psc\ sources in the association catalog with a given \simbad\ 
source type is shown in Table~\ref{table:simbadtypes}, for several different $P_{\rm id}$ cutoffs. 
The number of ``new'' associations---\bsc--\psc\ associations not present in \simbad---is also 
given for each source type. The new associations include 1464 high quality matches, 
679 medium quality matches, and 582 low quality matches with previously classified \psc\ sources. 
Of these new associations, only 39 high, 23 medium, and 66 low quality matches are with 
\bsc\ sources that are classified in \simbad. Thus, almost all the new associations with 
previously classified \psc\ sources provide probable new classifications for \bsc\ sources. 
For example, the high quality matches provide likely new \bsc\ classifications including 
125 (127 if ignoring previous \bsc\ classifications) double or multiple stars, 
173 (184) high proper motion stars, and 758 (772) stars.
\placetable{table:simbadtypes}

Also given in Table~\ref{table:simbadtypes} is the fraction of 
\psc\ sources of a given type that are high quality matches. 
This fraction illustrates the efficiency with which the method described in 
\S\ref{sec:method} associates high quality counterparts for a given source type 
(limited, of course, by any selection biases in \simbad). 
Stars, and other bright near-infrared point sources, 
are among the \ir\ counterparts with the largest fraction of high quality matches, 
while none of the 718 quasars or 306 Seyfert type galaxies have $P_{\rm id}>0.98$. 
This difference is most likely due to the definition of LR in the present work, which 
selects bright \ir\ sources.

\subsection{Color--Color Diagram}
\label{subsec:color}

A \bsc--\psc\ match has flux information both in an \xray\ band and in 
multiple \ir\ bands, and can be plotted with \xray--J color against 
J--K color. Such a color--color diagram is useful in studying the properties 
of the matches in the catalog. 

The \pspc\ count rate (0.1-2.4 keV) of each \bsc\ source is converted to an \xray\ flux (0.4-2.4 keV) 
using the HEASOFT tool {\tt pimms}, by assuming that the spectrum is a power-law of photon index 2. 
Two \xray\ fluxes are computed: the uncorrected \xray\ flux $F_{\rm
  X}$ (setting $N_{\rm H}=0$ in {\tt pimms}), 
and the absorption-corrected \xray\ flux $(F_{\rm X})_{\rm 0}$. The value of $N_{\rm H}$ used to 
calculate the absorption-corrected flux of each source is derived from the measurements of the 
Leiden/Argentine/Bonn Survey of Galactic HI \citep{kalberla05}, using
the HEASOFT tool {\tt nh}. 
The output of {\tt nh} chosen in this case is the weighted average of the data points from the survey closest to the 
direction of the \bsc\ source, and it should be noted that this corresponds to the total integrated 
column to the edge of the galaxy. Thus, the applied correction is systematically too large for 
sources within the Galaxy.

To calculate the extinction-corrected \ir\ magnitudes $(m_{\rm J})_{\rm 0}$ and $(m_{\rm K})_{\rm 0}$, 
it is necessary to relate the extinction in the two respective bands to the value of $N_{\rm H}$. 
Such a connection can be made by assuming that four separate relations between extinction 
corrections in various bands hold for all sources: 
The first relation \citep{predehl95}, 
\begin{equation}
N_{\rm H} = 1.79\tee{21}A_{\rm V} \, {\rm cm}^{-2} \, ,
\end{equation}
allows the V-band magnitude correction $A_{\rm V}$ to be found from the value of $N_{\rm H}$; 
the second relation \citep{schild77}, 
\begin{equation}
A_{\rm V} = 3.2\,  E_{\rm (B-V)} \, ,
\end{equation}
allows the B--V color correction $E_{\rm (B-V)}$ to be found from $A_{\rm V}$; 
and the final two relations \citep{fitzpatrick99},
\begin{equation}
A_{\rm J} = 0.829\,  E_{\rm (B-V)}
\end{equation}
and
\begin{equation}
A_{\rm K} = 0.355\,  E_{\rm (B-V)}\, ,
\end{equation}
allow the J- and K-band magnitude corrections $A_{\rm J}$ and $A_{\rm
K}$ to be found from $E_{\rm (B-V)}$. In the relation for $A_{\rm K}$,
the midpoint of the range (0.33-0.38) given for the prefactor is used.
Having computed the J-band magnitude extinction correction, the
corresponding extinction corrected flux is
\begin{equation}
\left(F_{\rm J}\right)_{\rm 0} = 5.082\tee{-7}\tee{-0.4(m_{\rm J}-A_{\rm J})}\, ,
\end{equation}
where the prefactor is the 'In-Band' zero-magnitude flux for the \twomass\ J-band \citep{cohen03}.

\begin{figure}[htb]
\plotone{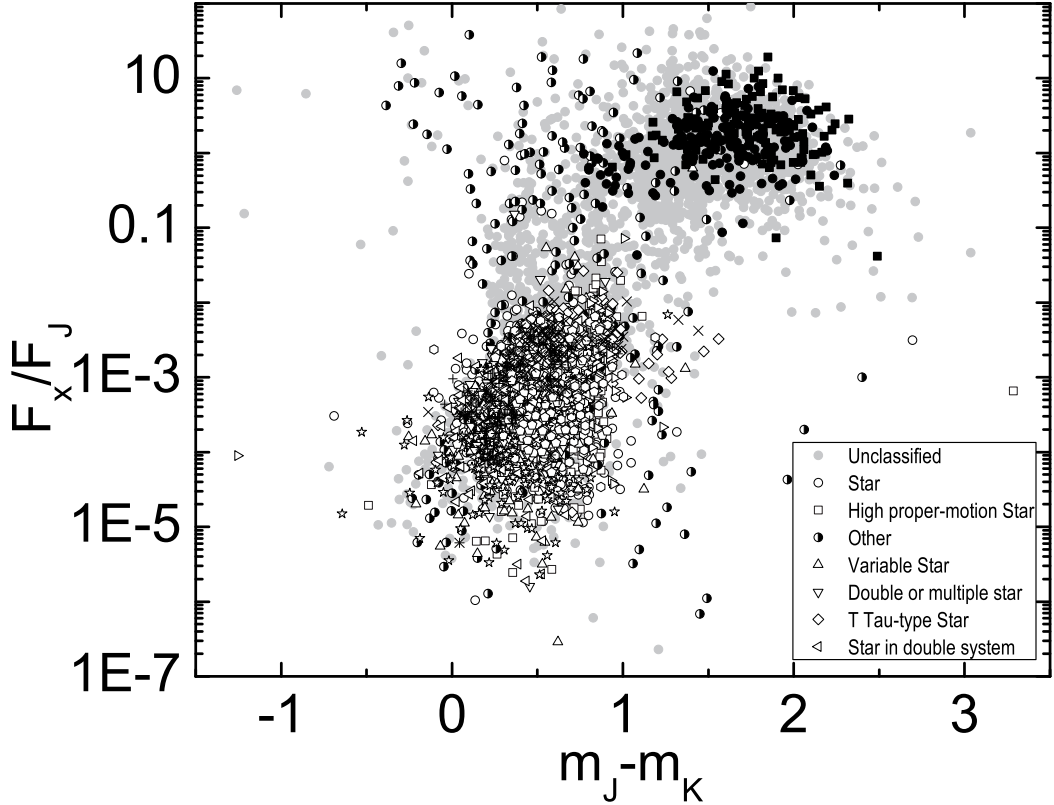}
\caption{
\xray\ to J-band flux ratio against J--K color of the $P_{\rm id}>0.5$ matches, 
where the point type reflects the source type listed in \simbad\ for the \psc\ source (\cf\ legend). 
Only source types for which there are at least 50 members are 
given a distinct point type, while the remaining classified sources are labeled as 'Other'. Sources 
not listed in \simbad, or listed as `X-ray source' or 'Infra-Red source', are labeled as 'Unclassified'. 
The source types are ordered by decreasing number of members (\cf\ Table~\ref{table:simbadtypes}), and the legend is split 
between this figure and Fig.~\ref{fig:FxFj} for legibility (the largest number of sources is 'Unclassified'). 
Two unclassified sources are omitted from the plot because they have 
$m_{\rm J}-m_{\rm K}<-1.5$ or $F_{\rm X}/F_{\rm J}>100$, and eight other sources are omitted 
because they do not have an $m_{\rm K}$ listed in the \psc.
Galaxies and coronally active stars appear to occupy distinct regions of the color--color diagram 
(\cf\ \S\ref{subsec:color}).
}
\label{fig:FxFjAbs}
\end{figure}

\begin{figure}[htb]
\plotone{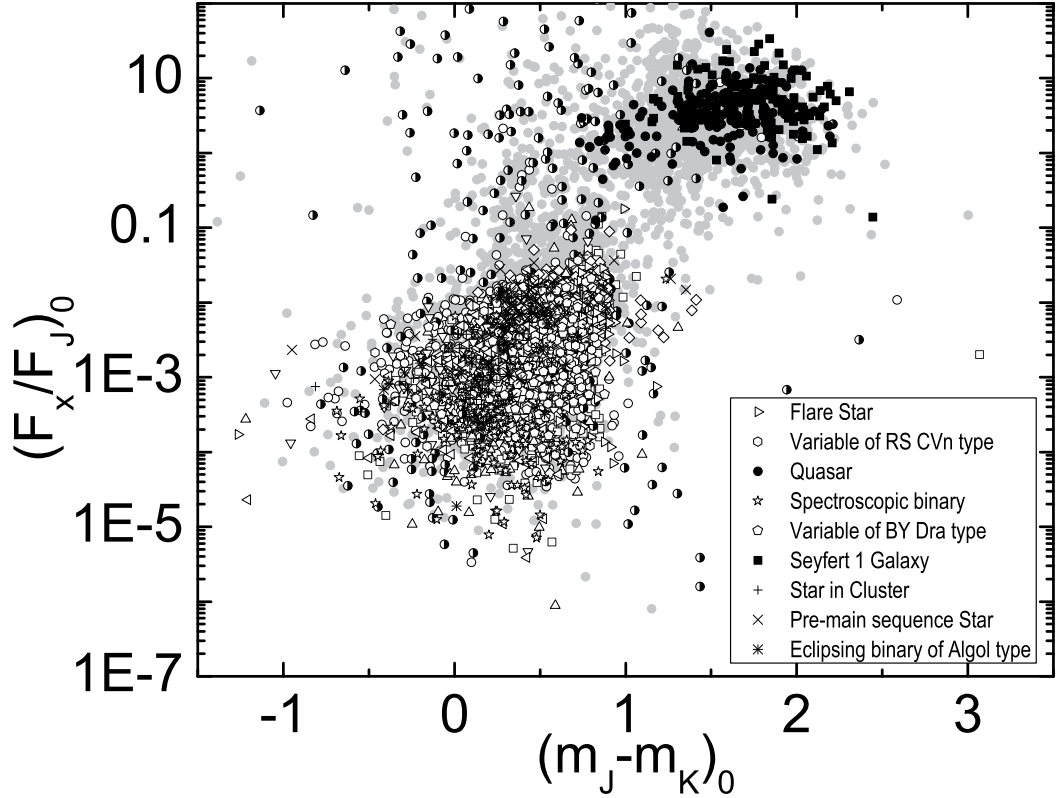}
\caption{
Same as Fig.~\ref{fig:FxFjAbs}, but corrected for \xray\ absorption and \ir\ extinction 
assuming the entire column of material to the edge of the galaxy lies between the earth and the 
source (\cf\ \S\ref{subsec:color}). This correction is expected to approximate the actual correction 
for extra-galactic sources, and is expected to be too large a correction for galactic sources.
Six unclassified sources with $(m_{\rm J}-m_{\rm K})_{\rm 0}<-1.5$ or 
$(F_{\rm X}/F_{\rm J})_{\rm 0}>100$ are omitted, in addition to the sources omitted in Fig.~\ref{fig:FxFjAbs}.
}
\label{fig:FxFj}
\end{figure}

The \xray\ to J-band flux ratio against J--K color of the $P_{\rm id}>0.5$ matches is shown in 
Fig.~\ref{fig:FxFjAbs} (uncorrected) and Fig.~\ref{fig:FxFj} (corrected for 
\xray\ absorption and \ir\ extinction). 
The source type listed in \simbad\ for the \psc\ source is shown for source types 
with $\geq$50 such $P_{\rm id}>0.5$ matches, 
while the remaining classified sources are labeled as 'Other'. 
Sources not listed in \simbad, or listed as `X-ray source' 
or 'Infra-Red source', are considered 'Unclassified'. 
The majority of classified sources are either galaxies or coronally active stars, where 'galaxies' 
include the source types 'Quasar' and 'Seyfert 1 Galaxy', while 'coronally active stars' include 
the remaining source types listed in the joint legend of Fig.~\ref{fig:FxFjAbs} and Fig.~\ref{fig:FxFj} 
(excluding 'Other' and 'Unclassified'). 
Galaxies and coronally active stars appear to occupy distinct regions of the color--color diagram: The galaxies 
have $m_{\rm J}-m_{\rm K}>(m_{\rm J}-m_{\rm K})_{\rm 0}>0.6$ and 
$(F_{\rm X}/F_{\rm J})_{\rm 0}>F_{\rm X}/F_{\rm J}>3\tee{-2}$, while almost all the coronally active stars have 
$(m_{\rm J}-m_{\rm K})_{\rm 0}<m_{\rm J}-m_{\rm K}<1.1$ and 
$F_{\rm X}/F_{\rm J}<(F_{\rm X}/F_{\rm J})_{\rm 0}<3\tee{-2}$. There are $3415$ unclassified 
$P_{\rm id}>0.5$ matches in the region occupied by coronally active stars, $1647$ in the 
region occupied by galaxies, and $442$ in the remainder of the color--color diagram.
\placefigure{fig:FxFjAbs}
\placefigure{fig:FxFj}

\begin{figure}[htb]
\plotone{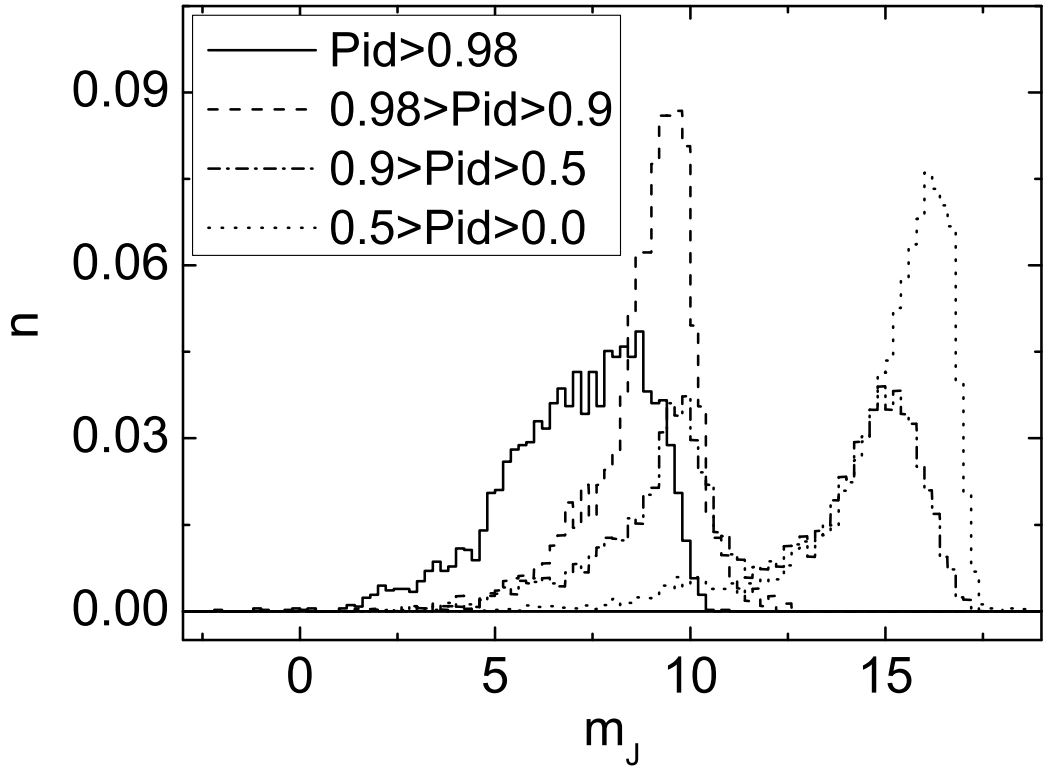}
\caption{Normalized number distributions in J-band magnitude $m_{\rm J}$ of \xray--\ir\ matches in the 
association catalog.
Almost all high and medium quality matches have $m_{\rm J}<12$. In the low quality and remaining matches, 
a population of fainter \ir\ sources is present. See \S\ref{subsec:catalog} for discussion.}
\label{fig:jmPid}
\end{figure}

The \xray\ flux of the bulk of the \bsc\ sources only spans a range of about one 
order of magnitude, so the large range in $F_{\rm X}/F_{\rm J}$ seen in Fig.~\ref{fig:FxFjAbs} 
is mainly due to the large range of J-band fluxes of sources in the \psc. 
Thus, the distribution of matches in $F_{\rm X}/F_{\rm J}$ is closely related to the 
distribution of matches in $m_{\rm J}$, which can be seen in Fig.~\ref{fig:jmPid}. It appears 
that the brighter population of sources seen in Fig.~\ref{fig:jmPid} consists mainly of 
coronally active stars.

To determine whether the upper limit on $F_{\rm X}/F_{\rm J}$ for coronally active stars is 
physical or due to selection effects (the chosen $LR$ favors bright J-band sources), a search for correlation between 
$F_{\rm X}$ and $F_{\rm J}$ was performed for the high quality matches for which the \psc\ source 
is of one of the stellar source types on the joint legend of Fig.~\ref{fig:FxFjAbs} and Fig.~\ref{fig:FxFj} 
(\ie\ not 'Unclassified', 
'Other', 'Quasar', or 'Seyfert 1 Galaxy'). Only high quality matches were considered to minimize the 
number of spurious matches, and to exaggerate any selection effects. A clear correlation was seen between $F_{\rm X}$ 
and $F_{\rm J}$ along the line corresponding to the upper limit on $F_{\rm X}/F_{\rm J}$, and the
probability that $F_{\rm X}$ and $F_{\rm J}$ are uncorrelated was found to be $\ee{-13}$ 
\citep[Spearman Rank--Order Correlation;][]{press92}. This correlation indicates a physical origin 
of the upper limit, as no direct dependence on $F_{\rm X}$ is included in $LR$, but a correlation 
between $F_{\rm X}$ and $F_{\rm J}$ could have been introduced through their mutual dependence on $N_{\rm H}$. 
However, the probability that $F_{\rm X}$ and $N_{\rm H}$ are uncorrelated is $6\tee{-2}$, and the 
probability that $F_{\rm J}$ and $N_{\rm H}$ are uncorrelated is 0.13 
\citep[Spearman Rank--Order Correlation;][]{press92}, so it is unlikely that a correlation between 
$F_{\rm X}$ and $F_{\rm J}$ could have been introduced through $N_{\rm H}$. 
The upper limit on $F_{\rm X}/F_{\rm J}$ thus appears to be physical in origin, 
and is most likely related to the saturation at $L_{\rm X}/L_{\rm bol}\approx\ee{-3}$ 
seen in rotation--activity diagrams for coronally active stars \citep{guedel04}.

\begin{figure}[htb]
\plotone{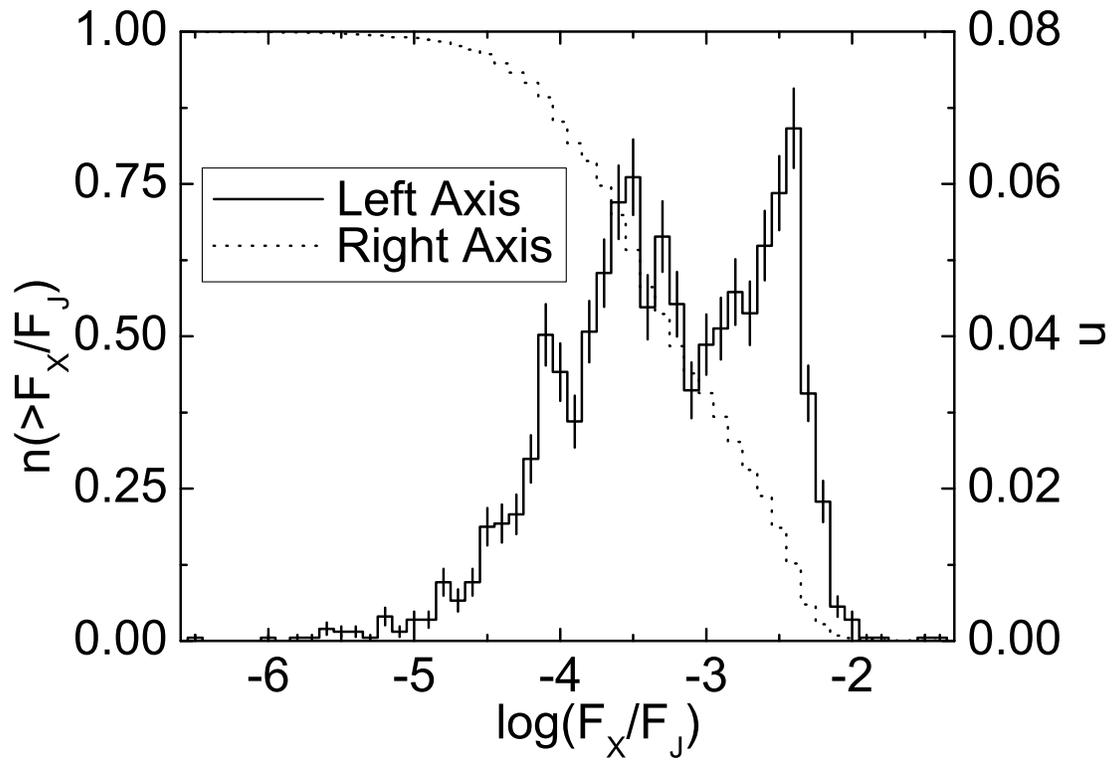}
\caption{[Right Axis] Normalized number distribution of the 2465 high quality matches that are 
coronally active stars against \xray\ to J-band flux ratio. See \S\ref{subsec:color} for discussion. 
[Left Axis] Cumulative number distribution of the same matches as for the right axis.
}
\label{fig:FxFjHistAbs}
\end{figure}

Fig.~\ref{fig:FxFjHistAbs} shows the differential and cumulative number distributions of 
high quality matches that are coronally active stars against $F_{\rm X}/F_{\rm J}$. 
The sharp cutoff at large $F_{\rm X}/F_{\rm J}$ is due to the previously discussed physical limit, 
while the scarcity of very bright J-band sources likely affects the shape of the distribution 
at small $F_{\rm X}/F_{\rm J}$. Since the \xray\ flux of the bulk of the \bsc\ only spans 
a range of about one order of magnitude, the cutoff due to the physical upper limit on $F_{\rm X}/F_{\rm J}$ 
for coronally active stars is the most likely cause of the similar cutoff seen in Fig.~\ref{fig:jmPid} 
around $m_{\rm J}=10$. Since the \psc\ is practically complete for $m_{\rm J}<15.8$ (\cf\ \S\ref{sec:intro}), 
all the coronally active stars in the \bsc\ should have counterparts in the \psc. 
The association catalog presented in \S\ref{subsec:catalog} is thus only 
confusion limited for J-band counterparts to coronally active stars, not flux limited.
\placefigure{fig:FxFjHistAbs}

The definition adopted for $LR$ favors 
bright J-band sources (\cf\ \S\ref{subsec:LR}). The association catalog therefore 
includes many high quality matches with bright J-band sources, like coronally active stars, 
but few matches with fainter \ir\ sources (\cf\ Fig.~\ref{fig:jmPid}). 
If high quality matches with fainter \ir\ sources are needed, the cross-association 
can be repeated with a different definition of $LR$. For example, if high quality matches with 
galaxies are needed, $LR$ could be defined to favor \ir\ objects with $(m_{\rm J}-m_{\rm K})_{\rm 0}>0.6$, 
in much the same way bright \ir\ objects were favored in the present cross-association.

\section{COMPARISON WITH OTHER PUBLISHED CROSS-ASSOCIATIONS}
\label{sec:comparison}

The cross-association method described in \S\ref{sec:method} is an adaptation to the
\psc\ of the method employed in previous work \xidp\ for the \usno\
catalog, so the results of the present work can be directly compared
to those of the \bsc--\usno\ cross-association. 
The quality $Q$ of the \usno\ association catalog is 0.652 \xidp, so it appears that 
the the fraction of the \bsc\ for which the present method can identify a \psc\ counterpart is 
is $7\pm 2\%$ larger than the fraction of the \bsc\ for which the previous work could 
identify a \usno\ counterpart. 
The number of counterparts found in both cross-associations is shown
in Table~\ref{table:cross-associations} for three different $P_{\rm
id}$ cut-offs. The present work finds 42\% more high quality matches
than the \bsc--\usno\ cross-association, but finds 9\% fewer $P_{\rm
id}>0.5$ matches. This difference in distribution of $P_{\rm id}$
values may be due to the use of a different passband (\ir) to
search for counterparts, which may be sensitive to
different source classes than the optical passband.
The difference may also be due to the changes made to $LR$; in
particular, the added $\rho$ dependence in $LR$ is expected to lead 
to higher $P_{\rm id}$ values for matches in regions with low \ir\
source density.  

Unlike the \bsc--\usno\ cross-association, the present work does not
consider binary matches.  A \bsc\ source associated with a pair of
\psc\ sources will typically have $P_{\rm no-id}\approx 0$, and appear
in the catalog as one of the $P_{\rm id}\la 0.5$ matches.

Of the high quality matches, 1418 are associations that are not listed
in the \simbad\ database, and that involve a \bsc\ source for which a
high quality match with a \usno\ optical source was not presented in
previous work. Excluding associations listed in \simbad\ and those
involving a \bsc\ source for which a high, medium, or low quality
\usno\ match has been previously been presented, the high quality
matches in the catalog include entirely new associations for 113 \bsc\
sources, the medium quality matches for 113 sources, and the low
quality matches for 808 sources.

Several cross-associations with other catalogs are described in the
\bsc\ paper \citep{voges99}, and these are compared to $P_{\rm id}$
based cross-associations in the \bsc--\usno\ cross-association \xidp.
In summary, accounting for source confusion, and not selecting
counterparts purely based on proximity, make $P_{\rm id}$ based
catalogs more useful than the previous cross-associations for
classifying \bsc\ sources.

Several of the cross-associations listed in
Table~\ref{table:cross-associations} rely on plausibility
arguments. One such example is a cross association with optical
sources \citep{zickgraf03}, where each optical source in the \bsc\
error circle was classified through spectroscopy, and the ``most
likely'' counterpart then assigned to the \xray\ source.  This
approach does not provide a $P_{\rm id}$ for each match, and relies on
previous knowledge of the properties of \xray\ emitting source
classes. It is therefore unlikely to identify new classes of \xray\
sources.

Machine learning algorithms can be employed to make the plausibility
method quantitative.  One application of machine learning algorithms
\citep{mcglynn04} used a training sample of previously classified
\xray\ sources to compute the probability that each \bsc\ source
belongs to a given source class. Machine learning algorithms remove
the need to specify the properties of a given class---the algorithms
infer class properties from the training sample---and the probability
that the \bsc\ source is of a certain class can be found. However,
only previously known classes can be used, and the number of objects
of a given class in the training sample limits the ability of the
algorithm to find similar objects.

\section{DISCUSSION AND CONCLUSIONS}
\label{sec:conclude}

The association catalog presented in \S\ref{subsec:catalog} lists each
\bsc\ source, the corresponding $P_{\rm no-id}$, the most likely \psc\
counterpart, and the $P_{\rm id}$ of the \bsc--\psc\ match. The
catalog includes 3853 high quality matches ($P_{\rm id}>0.98$), of
which $\leq39$ are expected to be chance associations with background
objects. Of the high quality matches, only 1921 are with a \bsc\
source for which a $P_{\rm id}>0.98$ \usno\ counterpart has previously
been identified \xidp. Thus, between the present work and the
\bsc--\usno\ cross-association, there are 4637 \bsc\ sources with a
$P_{\rm id}>0.98$ optical and/or \ir\ counterpart.
The present work thus represents a significant improvement over
previous cross-associations.

The purpose of uniquely associating \bsc\ and \psc\ sources is to
obtain J-, H-, and K-band fluxes for, and better localization of,
\bsc\ sources, to aid in classification.  Certain source classes
appear to be restricted to specific ranges of \xray--J color and J--K
color (\cf\ \S\ref{subsec:color}), so the \xray, J-, and K-band fluxes
of matches can be used to select interesting objects for precise
classification through follow-up observations.  Previous
classification of the proposed \psc\ counterpart in a match can also
provide \bsc\ classifications.  The present work presents likely new
classifications for 1425 previously unclassified \bsc\ sources, based
on the classification of the $P_{\rm id}>0.98$ \psc\ counterpart.

All the coronally active stars in the \bsc\ should have counterparts in 
the \psc\ (\cf\ \S\ref{subsec:color}), so unique association 
of these \bsc\ sources with \ir\ counterparts is confusion limited. 
For any coronally active stars in the \bsc\ that do not presently 
have a known \psc\ counterpart, better \xray\ localizations 
should allow the \psc\ counterpart to be found. 
Further, since the fraction of the \bsc\ sources that do not have counterparts 
in the \psc\ is $12\pm1\%$ (\cf\ \S\ref{subsec:fields}), there are $2200\pm200$ 
\bsc\ sources with no \ir\ counterparts in the \psc\ catalog. 
These \bsc\ sources are likely 
associated with \ir\ sources fainter than the \psc\ detection limit, 
and represent a sizeable discovery space for classes of \xray\ sources 
other than coronally active stars.

\bsc\ sources with $P_{\rm no-id}\approx 1$ have no likely \psc\
counterparts.  Some of these sources may be associated with an
extended near infra-red source not listed in the \psc, some may be
associated with \ir\ sources too far from the \bsc\ position to be
identified, and some may be \ir\ faint objects. The number of \bsc\
sources with $P_{\rm no-id}>0.98$ is 334, and these provide a sample
which can be searched for examples of \ir\ faint \xray\ sources like
isolated neutron stars.

\acknowledgments
The authors are grateful to the anonymous referee for careful reading of the manuscript, 
and for provocative questions that led to improvements to the text. 
C.~B.~H. gratefully acknowledges useful comments on the text by S.~E. Jackler. 
This publication makes use of data products from the Two Micron All Sky Survey, 
which is a joint project of the University of Massachusetts and the Infrared 
Processing and Analysis Center/California Institute of Technology, funded by 
the National Aeronautics and Space Administration and the National Science Foundation.
This research has made use of the SIMBAD database, operated at CDS, Strasbourg, France. 
This research has made use of NASA's Astrophysics Data System.
This work was supported by the NSERC Discovery Grants program.  

{\it Facilities:} \facility{ROSAT (PSPC)}, \facility{CTIO:2MASS}, \facility{FLWO:2MASS}

\clearpage
\bibliographystyle{apj_8}
\bibliography{references}

\clearpage

\clearpage
\begin{deluxetable}{rccc}
\rotate
\tablewidth{0pt}
\tabletypesize{\footnotesize}
\tablecaption{Published catalogs of counterparts for $N>500$ \bsc\ sources}
\tablehead{
\colhead{Reference} &
\colhead{Cross-association catalog\tablenotemark{a}} &
\colhead{$N$ $(N_{\rm bkg}=\% )$\tablenotemark{b}} &
\colhead{Criteria for association}
}
\startdata
\citet{huensch98} & Bright Star Catalogue & 580 $(<13=2.2\%)$ & $<90\arcsec $ from \xray\ source \\ 
\citet{bade98} & \bsc, $|b|>20\degr$ & 3847 (n/a) & Plausibility \\ 
 & Hamburg Quasar Survey & & \\
\citet{voges99} & HST-GSC & 9759 $(1358=13.92\%)$ & $<24\arcsec$ from \xray\ source \\ 
\citet{huensch99} & Third Catalogue of Nearby Stars & 783 $(<16=2\%)$ & $<90\arcsec $ from \xray\ source \\ 
\citet{bauer00} & \bsc\ $(>0.1\, {\rm counts\, s^{-1}})$,  & 1556 $(51=3.3\%)$ & $P_{\rm no-id}<0.5$ \\
 & $|b|>15\degr$, J2000 $\delta>-40\degr$ & & \\
 & NVSS & & \\
\citet{makarov00} & Tycho-2 & 6954 (n/a) & $<3\sigma_{\rm X}$ from \xray\ source \\
\citet{rutledge00} & USNO A-2, B objects & 2705 $(18=0.7\%)$ & $P_{\rm id}\geq 98\%$ \\
\citet{rutledge00} & USNO A-2, B objects & 5492 $(155=2.8\%)$ & $P_{\rm id}\geq 90\%$ \\
\citet{rutledge00} & USNO A-2, B objects & 11301 $(2034=18\%)$ & $P_{\rm id}\geq 50\%$ \\
\citet{schwope00} & \bsc\ $(>0.2\, {\rm counts\, s^{-1}})$, $|b|>30\degr$ & 2012 (n/a) & Plausibility \\
 & NED and SIMBAD & & \\
 & Optical CCD and spectroscopy & & \\
\citet{zimmermann01} & Catalogue of Principal Galaxies & 547 $(<66=12\%)$ & $<100\arcsec$ from \xray\ source \\
 & & & Visual Inspection \\
\citet{zickgraf03} & \bsc, $|b|>30\degr$ & 4380 (n/a) & Plausibility \\
 & Hamburg Quasar Survey & & \\
\citet{mcglynn04} & USNO-B & 6915 $(<1383=20\%)$ & $<30\arcsec$ from \xray\ source \\
 & NVSS and SUMSS & & Machine learning classification \\
\citet{torres06} & \bsc\ Southern Hemisphere & 1791 (n/a) & $<2.6\sigma_{\rm X}$ from \xray\ source \\
 & Hipparcos and Tycho-2 & & $B-V>0.6$ \\
Present work & \psc , J-band & 3853 $(39=1.0\%)$ & $P_{\rm id}> 98\%$ \\
Present work & \psc , J-band & 6133 $(140=2.3\%)$ & $P_{\rm id}> 90\%$ \\
Present work & \psc , J-band & 10286 $(1583=15.4\%)$ & $P_{\rm id}> 50\%$
\enddata
\tablenotetext{a}{The catalog that was searched for counterparts, 
and any restriction on which \bsc\ sources were considered.}
\tablenotetext{b}{The number of candiate counterparts presented, and the expected background contamination in 
number and as a percentage of the number of candidate counterparts.}
\label{table:cross-associations}
\end{deluxetable}

\clearpage
\begin{deluxetable}{ccccccc}
\centering
\tablewidth{0pt}
\tablecaption{Number of source and control fields in each of six different regions of the sky (\cf\ \S\ref{sec:results}), 
and the average $\rho$ for the fields in each region.}
\tablehead{
\colhead{Region} &
\colhead{Galactic Latitude} &
\colhead{Galactic Longitude} &
\colhead{$N_{\rm CON}$} &
\colhead{$\langle \rho_{\rm CON} \rangle$} &
\colhead{$N_{\rm SRC}$} &
\colhead{$\langle \rho_{\rm SRC} \rangle$} 
}
\startdata
Entire Sky & $|b|\leq 90\degr$ & $0\degr\leq l \leq 360\degr$ & 41192 & 3.17 & 18806 & 2.85 \\
1 & $50\degr<|b|$ & $0\degr\leq l \leq 360\degr$ & 9168 & 0.44 & 5194 & 0.44 \\
2 & $20\degr<|b|\leq 50\degr$ & $0\degr\leq l \leq 360\degr$ & 17270 & 1.06 & 7767 & 1.06 \\
3 & $|b|\leq 20\degr$ & $|l-180\degr| \leq 30\degr$ & 2455 & 3.27 & 1049 & 3.00 \\
4 & $|b|\leq 20\degr$ & $30\degr< |l-180\degr| \leq 90\degr$ & 4910 & 4.32 & 2036 & 3.98 \\
5 & $|b|\leq 20\degr$ & $90\degr< |l-180\degr| \leq 150\degr$ & 4932 & 9.12 & 1731 & 8.90 \\
6 & $|b|\leq 20\degr$ & $150\degr< |l-180\degr|$ & 2457 & 13.93 & 1029 & 14.22 \\
\enddata
\label{table:skypos}
\end{deluxetable}

\clearpage

\begin{deluxetable}{llccllcc}
\rotate
\centering
\tablewidth{0pt}
\tabletypesize{\footnotesize}
\tablecaption{Association catalog}
\tablehead{
\colhead{1RXS} &
\colhead{1RXS SIMBAD Type}  &
\colhead{PSPC c/s ($\sigma$)}  &
\colhead{$P_{no-id}$} &
\colhead{\twomass} &
\colhead{\twomass\ SIMBAD Type} &
\colhead{$m_J$} &
\colhead{$P_{id}$}
}
\startdata
J000000.0$-$392902 & X-ray source & 0.13 (0.03) & 0.654 & J00000425$-$3929005 & \nodata & 11.5 & 0.134\\
J000007.0$+$081653 & Galaxy & 0.19 (0.02) & 0.324 & J00000702$+$0816453 & \nodata & 13.9 & 0.670\\
J000010.0$-$633543 & Seyfert 1 Galaxy & 0.19 (0.03) & 0.449 & J00000981$-$6335383 & \nodata & 16.7 & 0.352\\
J000011.9$+$052318 & Seyfert 1 Galaxy & 0.26 (0.03) & 0.266 & J00001172$+$0523175 & \nodata & 14.4 & 0.680\\
J000012.6$+$014621 & X-ray source & 0.08 (0.02) & 0.008 & J00001217$+$0146173 & Double or multiple star & 8.3 & 0.991\\
J000013.5$+$575628 & Star & 0.12 (0.02) & 0.029 & J00001346$+$5756387 & \nodata & 9.8 & 0.960\\
J000019.5$-$261032 & X-ray source & 0.12 (0.02) & 0.329 & J00001916$-$2610283 & \nodata & 15.5 & 0.347\\
J000035.5$-$280553 & X-ray source & 0.07 (0.02) & 0.395 & J00003484$-$2805459 & \nodata & 12.9 & 0.474\\
J000038.4$+$794037 & Star & 0.10 (0.01) & 0.009 & J00004121$+$7940398 & Star & 8.9 & 0.966\\
J000042.5$+$621034 & Star & 0.16 (0.02) & 0.006 & J00004167$+$6210331 & Star & 6.1 & 0.989\\
\enddata
\tablecomments{The association catalog is available in machine readable format in the electronic edition of {\it The Astrophysical Journal}. A portion of the catalog is shown here for guidance regarding its form and content.}
\label{table:catalog}
\end{deluxetable}

\clearpage
\begin{deluxetable}{cccccc}
\centering
\tablewidth{0pt}
\tablecaption{Number of \psc\ sources in the association catalog that are 
classified in \simbad\ and/or listed as associated with the \bsc\ source in \simbad}
\tablehead{
\colhead{$P_{\rm id}$} &
\colhead{Total} &
\colhead{Associated} &
\colhead{Total} &
\colhead{Classified\tablenotemark{a}} &
\colhead{Unclassified\tablenotemark{b}} \\
 &
 &
\colhead{in \simbad?} &
 &
 &
}
\startdata
\multirow{2}{*}{$>0.98$} & \multirow{2}{*}{3853} & Yes & 1204  & 1186 & 18\\
 & & No & 2649 & 1464 & 1185 \\
\hline
\multirow{2}{*}{$>0.9$} & \multirow{2}{*}{6133} & Yes & 1579 & 1538 & 41 \\
 & & No & 4554 & 2143 & 2411 \\
\hline
\multirow{2}{*}{$>0.5$} & \multirow{2}{*}{10286} & Yes & 2122 & 2057 & 65 \\
 & & No & 8164 & 2725 & 5439 \\
\hline
\multirow{2}{*}{$>0$} & \multirow{2}{*}{18568} & Yes & 2609 & 2500 & 109 \\
 & & No & 15959 & 3386 & 12573 \\
\enddata
\tablenotetext{a}{Sources with a listed type other than 'X-ray source' or 'Infra-Red source'.}
\tablenotetext{b}{Sources not listed in \simbad, and sources listed as 'X-ray source' or 'Infra-Red source'.}
\label{table:SIMBAD}
\end{deluxetable}

\clearpage

\begin{deluxetable}{lrrrrrrrrrr}
\rotate
\centering
\tablewidth{0pt}
\tabletypesize{\scriptsize}
\tablecaption{Number of \psc\ sources in the association catalog listed in \simbad\ with a given source type}
\tablehead{
\colhead{Source type in \simbad} &
\multicolumn{2}{c}{$N(P_{id}>0.98)$}  &
\multicolumn{2}{c}{$N(P_{id}>0.9)$}  &
\multicolumn{2}{c}{$N(P_{id}>0.5)$}  &
\multicolumn{2}{c}{$N(P_{id}>0)$}  &
\multicolumn{2}{c}{$\frac{N(P_{id}>0.98)}{N_{Total}(P_{id}>0)}$\tablenotemark{b}} \\
 &
\colhead{Total} &
\colhead{New\tablenotemark{a}}  &
\colhead{Total} &
\colhead{New\tablenotemark{a}}  &
\colhead{Total} &
\colhead{New\tablenotemark{a}}  &
\colhead{Total} &
\colhead{New\tablenotemark{a}}  &
\colhead{Total} &
\colhead{New\tablenotemark{a}}  
}
\startdata
Active Galaxy Nucleus & 0 & 0 & 0 & 0 & 7 & 5 & 46 & 41 & 0.00 & 0.00 \\
Be Star & 4 & 2 & 5 & 3 & 5 & 3 & 5 & 3 & 0.80 & 0.40 \\
BL Lac - type object & 0 & 0 & 0 & 0 & 2 & 0 & 11 & 0 & 0.00 & 0.00 \\
Blazar & 0 & 0 & 0 & 0 & 0 & 0 & 1 & 0 & 0.00 & 0.00 \\
Blue object & 0 & 0 & 0 & 0 & 2 & 1 & 4 & 3 & 0.00 & 0.00 \\
Brown Dwarf ($M<0.08M_\sun$) & 0 & 0 & 0 & 0 & 1 & 0 & 1 & 0 & 0.00 & 0.00 \\
Carbon Star & 0 & 0 & 0 & 0 & 1 & 0 & 1 & 0 & 0.00 & 0.00 \\
Cataclysmic Var. AM Her type & 0 & 0 & 0 & 0 & 12 & 0 & 22 & 0 & 0.00 & 0.00 \\
Cataclysmic Var. DQ Her type & 1 & 0 & 1 & 0 & 8 & 0 & 9 & 0 & 0.11 & 0.00 \\
Cataclysmic Variable Star & 0 & 0 & 1 & 0 & 3 & 0 & 17 & 7 & 0.00 & 0.00 \\
Cepheid variable Star & 6 & 0 & 9 & 1 & 9 & 1 & 10 & 2 & 0.60 & 0.00 \\
Classical Cepheid (delta Cep type) & 1 & 1 & 1 & 1 & 1 & 1 & 1 & 1 & 1.00 & 1.00 \\
Double or multiple star & 177 & 127 & 219 & 159 & 242 & 177 & 252 & 184 & 0.70 & 0.50 \\
Dwarf Nova & 0 & 0 & 3 & 0 & 30 & 4 & 45 & 12 & 0.00 & 0.00 \\
Eclipsing binary & 7 & 2 & 7 & 2 & 9 & 3 & 9 & 3 & 0.78 & 0.22 \\
Eclipsing Binary Candidate & 1 & 0 & 1 & 0 & 1 & 0 & 1 & 0 & 1.00 & 0.00 \\
Eclipsing binary of Algol type & 37 & 18 & 43 & 20 & 51 & 26 & 51 & 26 & 0.73 & 0.35 \\
Eclipsing binary of beta Lyr type & 27 & 17 & 33 & 22 & 37 & 25 & 37 & 25 & 0.73 & 0.46 \\
Eclipsing binary of W UMa type & 30 & 4 & 40 & 6 & 45 & 8 & 45 & 8 & 0.67 & 0.09 \\
Ellipsoidal variable Star & 4 & 3 & 6 & 4 & 6 & 4 & 6 & 4 & 0.67 & 0.50 \\
Emission-line galaxy & 0 & 0 & 0 & 0 & 0 & 0 & 3 & 2 & 0.00 & 0.00 \\
Emission-line Star & 15 & 6 & 21 & 9 & 23 & 10 & 23 & 10 & 0.65 & 0.26 \\
Flare Star & 99 & 28 & 147 & 52 & 168 & 63 & 173 & 64 & 0.57 & 0.16 \\
Galaxy in Cluster of Galaxies & 0 & 0 & 0 & 0 & 0 & 0 & 1 & 1 & 0.00 & 0.00 \\
Globular Cluster & 0 & 0 & 0 & 0 & 1 & 0 & 2 & 0 & 0.00 & 0.00 \\
Herbig-Haro Object & 1 & 1 & 1 & 1 & 1 & 1 & 1 & 1 & 1.00 & 1.00 \\
High Mass X-ray Binary & 3 & 0 & 9 & 0 & 15 & 0 & 16 & 0 & 0.19 & 0.00 \\
High proper-motion Star & 286 & 184 & 392 & 264 & 477 & 337 & 493 & 353 & 0.58 & 0.37 \\
Low Mass X-ray Binary & 1 & 0 & 1 & 0 & 2 & 0 & 5 & 0 & 0.20 & 0.00 \\
Low-mass star ($M<1M_\sun$) & 0 & 0 & 1 & 0 & 4 & 3 & 10 & 9 & 0.00 & 0.00 \\
Nova & 0 & 0 & 1 & 0 & 2 & 0 & 2 & 0 & 0.00 & 0.00 \\
Nova-like Star & 2 & 0 & 3 & 0 & 6 & 0 & 10 & 2 & 0.20 & 0.00 \\
Object of unknown nature & 0 & 0 & 0 & 0 & 0 & 0 & 2 & 2 & 0.00 & 0.00 \\
Planetary Nebula & 0 & 0 & 1 & 0 & 1 & 0 & 1 & 0 & 0.00 & 0.00 \\
Possible Quasar & 0 & 0 & 0 & 0 & 0 & 0 & 6 & 0 & 0.00 & 0.00 \\
Pre-main sequence Star (optically detected) & 15 & 1 & 44 & 3 & 64 & 10 & 71 & 10 & 0.21 & 0.01 \\
Pulsating variable Star & 9 & 3 & 11 & 4 & 11 & 4 & 12 & 5 & 0.75 & 0.25 \\
Quasar & 0 & 0 & 1 & 0 & 161 & 65 & 718 & 401 & 0.00 & 0.00 \\
Rotationally variable Star & 12 & 4 & 12 & 4 & 13 & 4 & 14 & 4 & 0.86 & 0.29 \\
Semi-regular pulsating Star & 15 & 8 & 15 & 8 & 16 & 8 & 16 & 8 & 0.94 & 0.50 \\
Seyfert 1 Galaxy & 0 & 0 & 1 & 0 & 135 & 10 & 304 & 61 & 0.00 & 0.00 \\
Seyfert Galaxy & 0 & 0 & 0 & 0 & 1 & 0 & 2 & 0 & 0.00 & 0.00 \\
Spectroscopic binary & 163 & 86 & 185 & 100 & 205 & 111 & 207 & 113 & 0.79 & 0.42 \\
Star & 1099 & 772 & 1586 & 1190 & 1875 & 1439 & 2024 & 1580 & 0.54 & 0.38 \\
Star in Association & 2 & 1 & 7 & 3 & 10 & 5 & 12 & 7 & 0.17 & 0.08 \\
Star in Cluster & 41 & 18 & 60 & 29 & 65 & 30 & 71 & 35 & 0.58 & 0.25 \\
Star in double system & 67 & 44 & 104 & 77 & 208 & 155 & 222 & 168 & 0.30 & 0.20 \\
Star in Nebula & 0 & 0 & 0 & 0 & 1 & 1 & 2 & 2 & 0.00 & 0.00 \\
SuperNova & 0 & 0 & 0 & 0 & 1 & 1 & 1 & 1 & 0.00 & 0.00 \\
SuperNova Remnant & 0 & 0 & 0 & 0 & 0 & 0 & 1 & 1 & 0.00 & 0.00 \\
Symbiotic Star & 4 & 3 & 5 & 4 & 7 & 4 & 7 & 4 & 0.57 & 0.43 \\
T Tau-type Star & 94 & 2 & 185 & 12 & 238 & 15 & 247 & 17 & 0.38 & 0.01 \\
UV-emission source & 0 & 0 & 0 & 0 & 0 & 0 & 2 & 2 & 0.00 & 0.00 \\
Variable of BY Dra type & 79 & 35 & 91 & 41 & 107 & 49 & 109 & 51 & 0.72 & 0.32 \\
Variable of RS CVn type & 110 & 16 & 123 & 19 & 136 & 24 & 136 & 24 & 0.81 & 0.12 \\
Variable Star & 198 & 61 & 249 & 80 & 278 & 87 & 283 & 90 & 0.70 & 0.22 \\
Variable Star of alpha2 CVn type & 2 & 0 & 2 & 0 & 2 & 0 & 3 & 1 & 0.67 & 0.00 \\
Variable Star of beta Cep type & 5 & 2 & 7 & 3 & 7 & 3 & 7 & 3 & 0.71 & 0.29 \\
Variable Star of delta Sct type & 7 & 5 & 7 & 5 & 8 & 6 & 8 & 6 & 0.88 & 0.62 \\
Variable Star of gamma Dor type & 6 & 2 & 6 & 2 & 7 & 3 & 7 & 3 & 0.86 & 0.29 \\
Variable Star of irregular type & 5 & 5 & 6 & 5 & 6 & 5 & 6 & 5 & 0.83 & 0.83 \\
Variable Star of Mira Cet type & 1 & 0 & 2 & 1 & 2 & 1 & 2 & 1 & 0.50 & 0.00 \\
Variable Star of Orion Type & 3 & 2 & 8 & 4 & 14 & 7 & 19 & 9 & 0.16 & 0.11 \\
Variable Star of RR Lyr type & 0 & 0 & 2 & 1 & 3 & 1 & 4 & 2 & 0.00 & 0.00 \\
Variable Star of W Vir type & 1 & 0 & 1 & 0 & 1 & 0 & 1 & 0 & 1.00 & 0.00 \\
Variable Star with rapid variations & 1 & 0 & 1 & 0 & 1 & 0 & 1 & 0 & 1.00 & 0.00 \\
White Dwarf & 4 & 1 & 5 & 1 & 18 & 2 & 31 & 5 & 0.13 & 0.03 \\
Wolf-Rayet Star & 3 & 0 & 5 & 1 & 5 & 1 & 5 & 1 & 0.60 & 0.00 \\
X-ray Binary & 1 & 0 & 1 & 0 & 1 & 0 & 4 & 0 & 0.25 & 0.00 \\
Young Stellar Object & 1 & 0 & 3 & 2 & 3 & 2 & 4 & 2 & 0.25 & 0.00 \\
Young Stellar Object Candidate & 0 & 0 & 0 & 0 & 0 & 0 & 1 & 1 & 0.00 & 0.00 \\
\tableline
All Classes & 2650 & 1464 & 3681 & 2143 & 4782 & 2725 & 5886 & 3386 & 0.45 & 0.25 \\
\enddata
\tablenotetext{a}{\psc\ counterparts that are not associated with the \bsc\ source in \simbad.}
\tablenotetext{b}{Fraction of \psc\ counterparts of a given class that are high quality matches.}
\label{table:simbadtypes}
\end{deluxetable}

\end{document}